\documentclass[manuscript]{aastex}
\usepackage{graphicx}
\usepackage{natbib}
\usepackage{amssymb}

\slugcomment{To appear in Ap. J.}

\begin{document}

\title{{\it Spitzer}-IRAC survey of molecular jets in Vela-D}

\author{T. Giannini\altaffilmark{1}, D. Lorenzetti\altaffilmark{1},  M. De Luca\altaffilmark{2}, F. Strafella\altaffilmark{3}, D. Elia\altaffilmark{4}, B. Maiolo\altaffilmark{3}, M. Marengo\altaffilmark{5}, Y. Maruccia\altaffilmark{3}, F. Massi\altaffilmark{6}, B. Nisini\altaffilmark{1}, L. Olmi\altaffilmark{6,7}, A. Salama\altaffilmark{1}, H. A. Smith\altaffilmark{8}
}
\altaffiltext{1}{INAF - Osservatorio Astronomico di Roma, via Frascati 33, 00040 Monte Porzio, Italy}
\altaffiltext{2}{LERMA - LRA, UMR 8112 du CNRS, Observatoire de Paris, \'Ecole Normale Sup\'erieure, UPMC UCP, 24 rue Lhomond, 75231 Paris Cedex 05, France}
\altaffiltext{3}{Dipartimento di Matematica e Fisica - Universit\`a del Salento, CP 193, 73100, Lecce, Italy}
\altaffiltext{4}{INAF - Istituto di Astrofisica e Planetologia Spaziali, via Fosso del Cavaliere 100, 00133 Roma, Italy}
\altaffiltext{5}{Department of Physics and Astronomy, Iowa State University, Ames, IA 50011, USA}\altaffiltext{6}{Harvard-Smithsonian Center for Astrophysics, Cambridge, MA, USA}
\altaffiltext{6}{INAF - Osservatorio Astrofisico di Arcetri, Largo E. Fermi 5, 50125 Firenze, Italy}
\altaffiltext{7}{University of Puerto Rico, Rio Piedras Campus, Physics Dept., Box 23343, UPR station, San Juan, Puerto Rico (USA)}
\altaffiltext{8}{Harvard-Smithsonian Center for Astrophysics, Cambridge, MA, USA}
\begin{abstract}
We present a survey of H$_2$ jets from young protostars in the Vela-D molecular cloud (VMR-D), based on \emph{Spitzer}-IRAC data between 3.6\,$\mu$m and 8.0\,$\mu$m. Our search has led to the identification of 15 jets (2 new discoveries) and about 70 well aligned knots within 1.2 deg$^2$. We compare the IRAC maps with observations of the H$_2$ 1-0 S(1) line at 2.12\,$\mu$m, with a \emph{Spitzer}-MIPS map at 24\,$\mu$m and 70\,$\mu$m, and with a map of the dust continuum emission at 1.2\,mm. From such a comparison we find a tight association between molecular jets and dust peaks.\\
The jet candidate exciting sources have been searched for in the published catalog of the Young Stellar Objects of VMR-D. In particular, we searched for all the
sources of Class II or (preferentially) earlier which are located close to the jet center and aligned with it. Furthermore, the association between jet and exciting source was validated by estimating the differential extinction between the jet opposite lobes. We are able to find a best-candidate exciting source in all but two jets, for which two alternative candidates are 
given. Four exciting sources are not (or very barely) observed at wavelengths shorter than 24\,$\mu$m, suggesting they are very young protostars. Three of them are also associated with the most compact jets (projected length $\la$ 0.1 pc).\\
The exciting source Spectral Energy Distributions (SEDs) have been constructed and modeled by means of all the available photometric data between 1.2\,$\mu$m and 1.2\,mm. From SEDs fits we derive the main source parameters, which indicate that 
most of them are low-mass protostars.\\
A significant correlation is found between the projected jet length and the [24] - [70] color, which is consistent with an evolutionary scenario according to which shorter jets are associated with younger sources. A rough correlation is found between IRAC line cooling and exciting source bolometric luminosity, in agreement with the previous literature. 
The emerging trend suggests that mass loss and mass accretion are tightly related phenomena and that both decrease with time. 
\end{abstract}



\section{Introduction}\label{sec:sec1}
Supersonic flows observed as collimated jets are frequently associated with young protostars of different masses and evolutionary
stages. Jets represent an essential tool by which the driving source removes its excess angular momentum, allowing the collapse to go on.
Their interaction with the ambient medium occurs via excitation of atoms and ions, whose cooling lines represent the way through
which jets are seen. Such cooling lines spread over a wide frequency range from optical-UV up to far-IR. In particular, molecular hydrogen imaging
of the $v$ = 1-0 S(1) line at 2.12 $\mu$m is extensively used for identifying knots of molecular emission, while the spectroscopy of
the H$_2$ ro-vibrational lines lying in the near-IR range (1-5\,$\mu$m) is largely effective both to probe the molecular gas at
thousands Kelvin and to infer the main excitation mechanism(s) (fluorescence, C-ontinous or J-ump shocks, see e.g.
Black $\&$ van Dishoeck 1987; Kaufman \& Neufeld 1996; Hollenbach $\&$ McKee 1989; Smith 1995).  

Statistically, jet images are powerful in correlating the properties of the young population of the parental cloud (e.g. age, degree of clustering,
Initial Luminosity Function - ILF) with their common features  (e.g velocities, dynamical ages, collimation and position angles). Such correlations 
are undoubtedly fundamental for investigating how the cloud parameters affect the onset, the properties and the
morphology of the jets and how these latter influence the cloud itself. 

During the last two decades, hundreds of individual Young Stellar Objects (YSOs) have been imaged with near-infrared narrow band filters
(H$_2$ and [Fe II]), although extensive surveys able to cover square degree scale star forming regions are not so numerous (for some
exceptions see eg. Stanke et al. 1998, Davis et al. 2008, 2009). In this sense the digital catalog of Molecular Hydrogen Objects (MHO, Davis et al. 2010) undoubtedly represents a relevant tool of investigation.

The {\it InfraRed Array Camera} (IRAC, Fazio et al. 2004) on board of the {\it Spitzer Space Telescope} (Werner et al. 2004)
offers a unique chance to carry out large field searches of both jets and their obscured exciting sources (e.g. Cyganowski et al. 2011),  Indeed, IRAC band-passes (between 3.6\,$\mu$m and 8.0\,$\mu$m)
contain numerous H$_2$ emission lines (see Fig.1 of Smith \& Rosen 2005) and are therefore suited to find jets and to study their properties, as testified by an increasing 
number of observational works (e.g. Teixeira et al. 2008; Smith et al. 2006; Velusamy et al. 2007; Neufeld \& Yuan 2008, Takami et al. 2011, Ybarra et al. 2010, Ohlendorf et al. 2012, Lee et al. 2012). Such an observational capability led to the
implementation of detailed interpretative tools (Smith \& Rosen 2005; Neufeld \& Yuan 2008; Ybarra \& Lada 2009 (YL09); Takami et al. 2010) able to analyze IRAC maps of molecular flows. 
However, while the mere jet identification is a straightforward task, the derivation of the physical conditions from such photometric maps has to be done with some caution (Noriega-Crespo et al. 2004, Neufeld \& Yuan 2008). Indeed the H$_2$ emission lines inside the IRAC bands come from vibrational levels corresponding to very different excitation temperatures (from 800 K to 4000 K), and contamination from PAH, CO and atomic lines can be significant (Takami et al. 2010). 

In addition to allow jets identification, mid-infrared IRAC images have a major role in finding the embedded exciting sources that could have escaped detection at shorter wavelengths. Otherwise, IRAC observations allow to
probe the continuum emission of already identified exciting sources, but in a frequency regime where most of their energy is likely emitted. Particularly suited are the IRAC surveys of star
forming regions for which a census of the young stellar population has been already taken, since they offer an immediate correlation between the identified jets and the nature of the exciting sources. A well suited test-bed
is the cloud D of the Vela Molecular Ridge (hereinafter VMR-D, Murphy \& May 1991), an active site of star formation in the Galactic plane showing a rich phenomenology
associated with star formation processes, such as outflows (Wouterloot \& Brand 1999; Elia et al. 2007), jets (Lorenzetti et al. 2002b; Giannini et al. 2005), and evidences of clusters as well
as isolated protostars (Massi et al. 2000, 2003, Giannini et al. 2007, Strafella et al. 2010, hereafter SEC10). 

In the present paper we analyze the IRAC maps of VMR-D in order to : {\it i}) obtain an IRAC-based census
of the protostellar jets in VMR-D, {\it ii}) compare the jets found in the IRAC images with those we have already detected in the H$_2$ 1-0S(1) at 2.12\,$\mu$m and validate (or not) the applied IRAC
method; {\it iii}) identify new exciting sources or provide the mid-IR photometry for those already known; {\it iv}) derive the morphology of the jets, making use also of data 
in other bands, and {\it v}) correlate the properties of the jets with those of their exciting sources. 

The paper has the following structure: we first briefly present the data (Sect.\,2) and explain how jets are
identified and analyzed (Sect.\,3). In Sect.\,4 we search for the jet exciting sources and present their photometry. In Sect.\,5 we discuss our results, that are briefly summarized in Sect.\,6.
Finally, we give in Appendix\,\ref{sec:appendix} a brief description of each detected jet. 


\section{Observational data}
\subsection{IRAC, MIPS and millimeter data}
We have observed VMR-D as part of the IRAC-\emph{Spitzer} Cycle 3 GTO (PID 30335, PI: G. Fazio). The observations cover a region of about 1.2 square degree delimited in longitude by 264$^{\circ} 29^{\prime} \la {\it l} \la 263^{\circ} 00^{\prime}$ and in latitude by 0$^{\circ} 42^{\prime} \la {\it b} \la −0^{\circ}
50^{\prime}$. We have presented the photometric results in SEC10, who describe in detail the observational strategy, source extraction and photometry. This work resulted in a catalog, which we use in the present paper to identify the jet exciting sources (ES). In the same catalog it is also listed the MIPS-\emph{Spitzer} photometry at 24\,$\mu$m and 70\,$\mu$m (Giannini et al. 2007, SEC10) as well the 1.2 mm photometry (Massi et al. 2007, referred also as SIMBA photometry), which are used here to construct the SED of the exciting sources. 

\subsection{H$_2$ 2.12\,$\mu$m images}
To compare the \emph{Spitzer} maps with the H$_2$ emission in the near-infrared, we use images of the H$_2$ 1-0\,S(1) ro-vibrational transition at 2.12\,$\mu$m we obtained on December 2001 with SofI (Lidmann et al. 2006) at ESO-NTT (La Silla, Chile). These images cover the peaks of the dust continuum emission found by Massi et al. (2007) and have been already searched for evidences of shocked emission by Lorenzetti et al. (2002b) and  De Luca et al. (2007). Out of 18 H$_2$ SofI fields, we have detected shocked emission in the
11 fields displayed in Figure\,\ref{all:fig}, which are overlaid on the IRAC image of VMR-D at 4.5\,$\mu$m (left panel). In the same Figure, right panel, the dust emission image at 1.2\,mm is depicted, as well. In one case (jet\,\# 15) the 2.12\,$\mu$m image has been obtained with ISAAC (Cuby et al. 2004) at ESO-VLT (Paranal, Chile), and reported in Giannini et al. (2005).

\subsection{Complementary data}
\subsubsection{Public surveys data}
In order to supplement the photometry of the ES of the jets in the wavelength ranges adjacent to those probed with 
IRAC/MIPS, we used the 2MASS (Skrutskie et al. 2006) photometry in the $J$, $H$ and $K_s$ bands, and the recently delivered data 
of the \emph{Wide-field Infrared Survey Explorer} (WISE, Wright et al. 2010) taken at 
3.4\,$\mu$m, 4.6\,$\mu$m, 12\,$\mu$m, and 22\,$\mu$m. Finally, data obtained at 250\,$\mu$m, 350\,$\mu$m, and 500\,$\mu$m with the Balloon-borne 
Large-Aperture Submillimeter Telescope (BLAST, Pascale et al. 2008) and compiled by Netterfield et al. (2009) and Olmi et al. (2010) have 
also been considered in the SED's building.  

\subsubsection{CO APEX data}
$^{12}$CO(3-2) maps of four jets (namely \# 1, 2, 3, 5) were obtained from September to December 2006 with the 
APEX-2A heterodyne receiver, mounted on the APEX telescope (G\"usten et al. 2006) at Llano Chajnantor, Chile. 
The maps are typically 40 arcseconds wide with a velocity channel width of 0.5 km s$^{-1}$. The analysis of these maps will
be presented in a forthcoming paper, while here they will be only marginally used in Sect.\ref{sec:jetext} for a comparison with the IRAC maps.


\section{Jet analysis}
\begin{figure}
\includegraphics[angle=90,width=9.3cm]{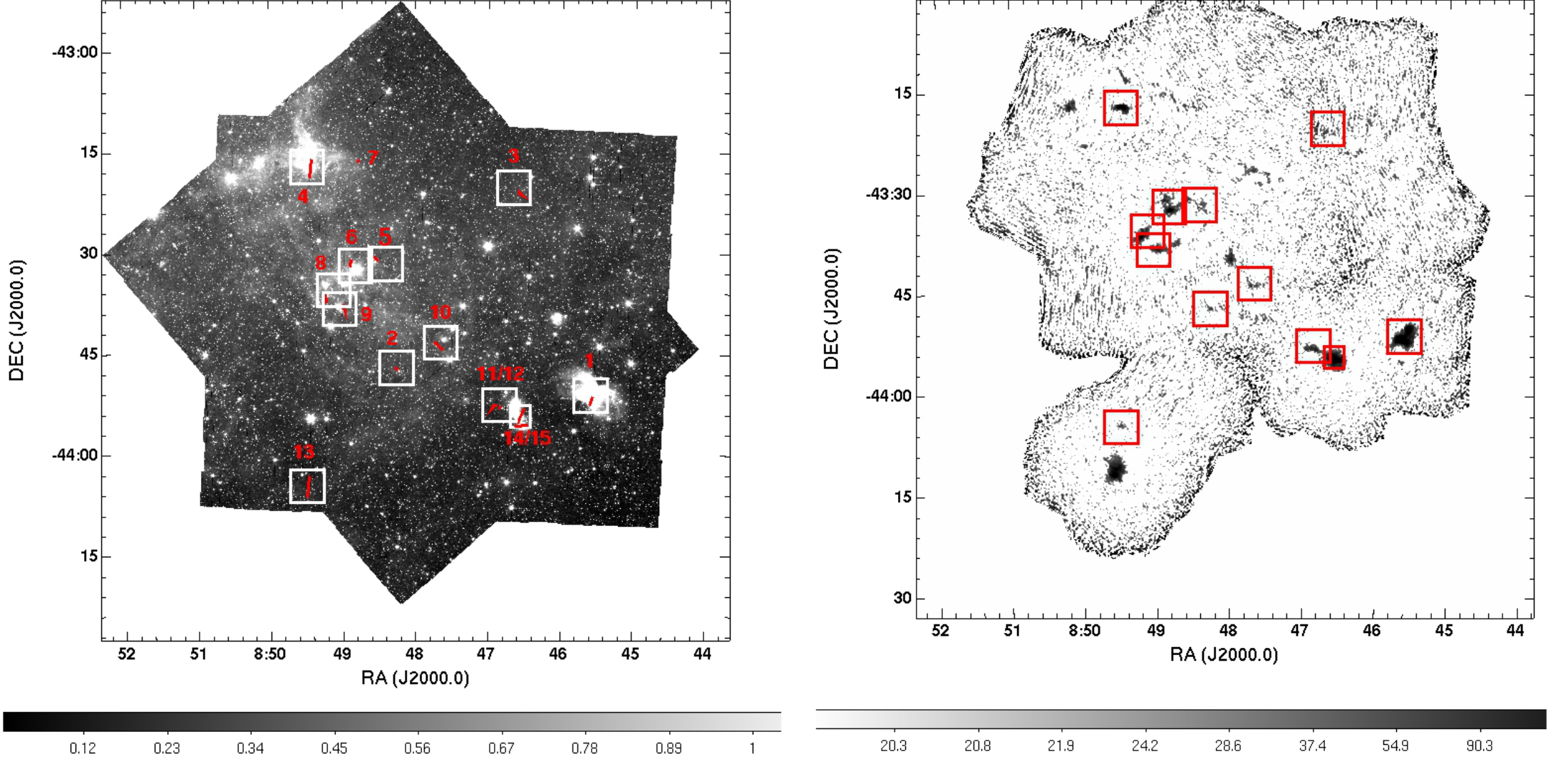}
\caption{{\it Left panel} - IRAC image of VMR-D at 4.5\,$\mu$m (relative intensity is given in the bottom bar). White boxes represent the fields imaged with SofI at 2.12\,$\mu$m (ISAAC for jet\,\# 15) where jets have been discovered (represented with red lines). Numbered labels refer to the jet identification numbers listed in Table\,\ref{tab:tab1}.
{\it Right panel} - Map of the dust continuum at 1.2 mm (adapted from Massi et al. 2007). The SofI fields (here shown in red) cover most of the brightest peaks. Intensity is given in the bottom bar in units of mJy/beam. \label{all:fig}}
\end{figure}

\begin{figure}
\includegraphics[angle=0,width=12cm]{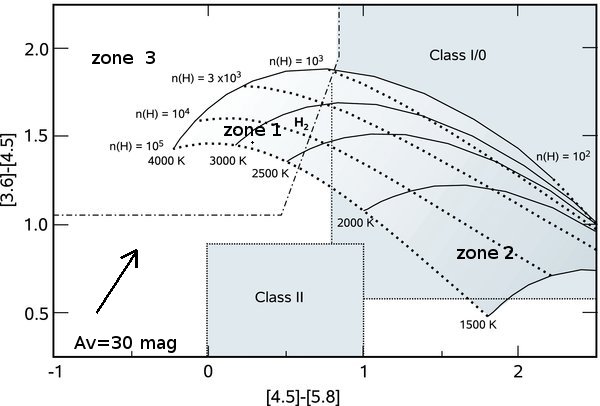}
\caption{Adapted from Figure 1 of YL09. IRAC [3.6] - [4.5] vs. [4.5] - [5.8] two colors plot indicating the region occupied by
shocked H$_2$ emission. Constant density (dotted) and temperature (solid) lines are indicated. Inside the diagram three regions are
defined : {\it zone 1 -} region encompassed by curves with 10$^3$ cm$^{-3} \la n($H$) \la 10^5$ cm$^{-3}$ and 2300 K $\la T \la$ 4000 K and
having [4.5] - [5.8] $\la$ 0.8 mag; {\it zone 2 -} region possibly affected by contamination of Class 0/I objects: i.e. within the curves with 
10$^2$ cm$^{-3} \la n$(\rm{H}$) \la 10^5$ cm$^{-3}$ and 1500 K $\la T \la$ 2300 K and with [4.5] - [5.8] $\ga$ 0.8; {\it zone 3 -} region empirically determined by Gutermuth et al. (2008) to contain outflows (dashed-dotted line, i.e. roughly the portion with [4.5] - [5.8] $\la$ 0.8 mag  and [3.6] - [4.5] $\ge$ 1.05 mag). The reddening vector,
corresponding to A$_V$=30 mag, is depicted as well. \label{YL_zones:fig}}
\end{figure}

\begin{figure}
\includegraphics[angle=0,scale=.30]{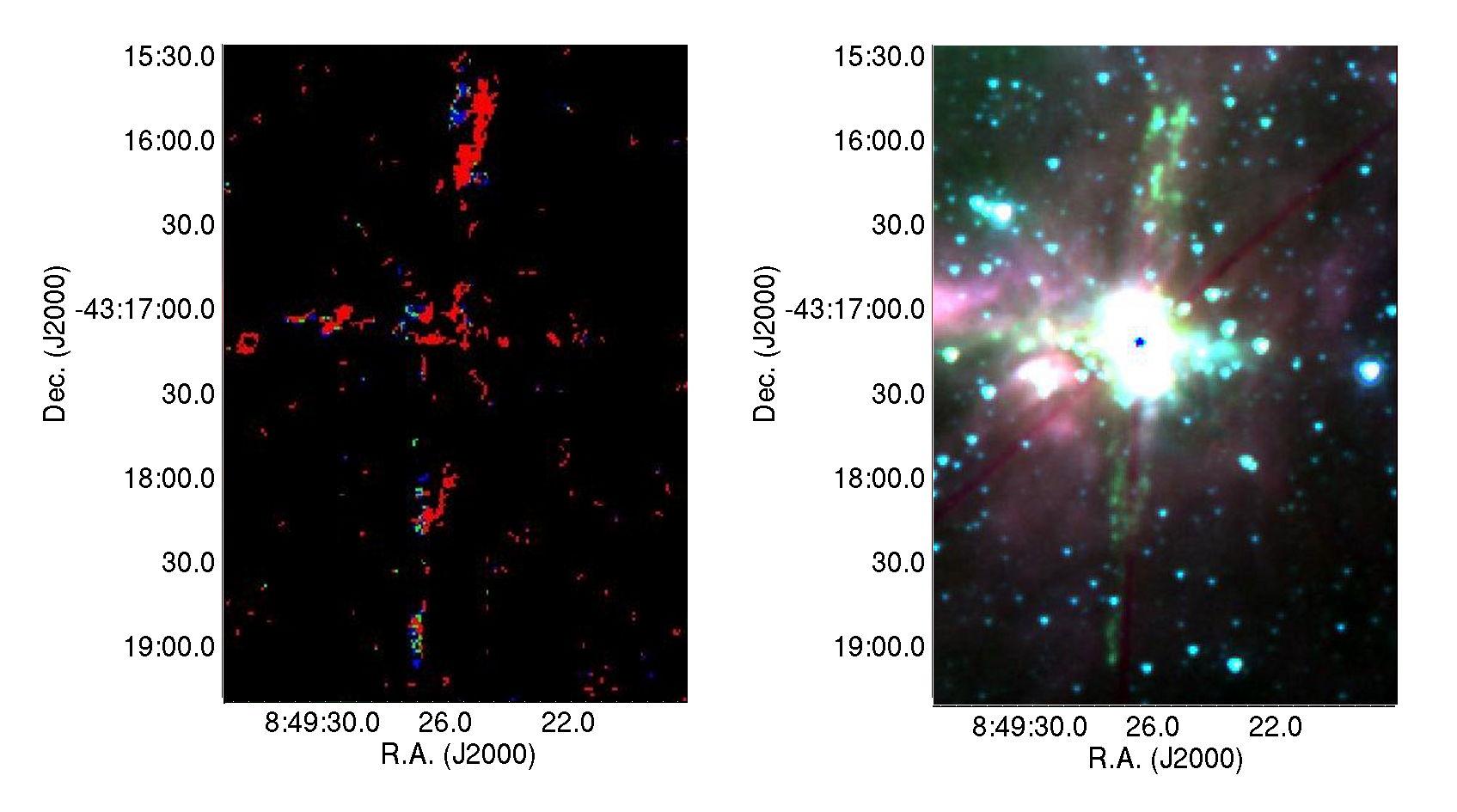} 
\caption{{\it Left panel} - Shock-mask around the location of jet\,\# 4 of all the pixels having IRAC
colors compatible with shocked H$_2$. Colors (green, red and blue) are those corresponding to the three zones (1,2,3, respectively) depicted in
Figure~\ref{YL_zones:fig}. {\it Right panel} - Same portion of the sky depicted in a three-color map obtained by combining the IRAC images in the three bands. The blue
channel is used at 3.6\,$\mu$m, the green one at 4.5\,$\mu$m, and the red one at 5.8 $\mu$m. Note that, while the shock-mask 
evidences two possible jets in orthogonal directions, the three-color image confirms 
the presence only of the jet in the vertical direction (see text for more details).\label{IRcolcol:fig}}
\end{figure}


\subsection{Jet searching procedure}
YL09 have shown how the analysis of \emph{Spitzer} data can be used to discover 
and characterize the emission from protostellar outflows. Such a 
characterization by means of the
IRAC color-color plot relies on the presence of strong (0-0) and (1-0) 
H$_2$ emission features in the IRAC bands (Neufeld \& Yuan 2008), and 
makes it possible to define, on a [3.6] - [4.5] vs.
[4.5] - [5.8] IRAC diagram, the region occupied by shocked H$_2$ as a 
function of the gas kinetic temperature and hydrogen density, in the range 
1500 K $\le T \le$ 4000 K and 10$^2$  cm$^{-3}$ $\le n$(\rm{H})$\le$ 10$^5$ 
cm$^{-3}$, respectively.

In Figure~\ref{YL_zones:fig} (adapted from Figure 1 of YL09), we show three regions (jet {\it locus}) identified by YL09
to contain jets: (1) the region corresponding to jets characterized by sufficiently high temperature 
and hydrogen density, that are unambiguously distinguished from stellar objects; i.e. the region 
encompassed by curves with 10$^3$ cm$^{-3} \la n($H$) \la 10^5$ cm$^{-3}$ and 2300 K $\la T \la$ 4000 K and 
having [4.5] - [5.8] $\la$ 0.8 mag, zone 1; (2) the region possibly affected by contamination of Class 0/I objects:
i.e. within the curves with 10$^2$ cm$^{-3} \la n($H$) \la 10^5$ cm$^{-3}$ and 1500 K $\la T \la$ 2300 K and 
with [4.5] - [5.8] $\ga$ 0.8 mag, zone 2; (3) the region empirically determined by Gutermuth et al. (2008)
to contain outflow, i.e. roughly the portion with [4.5] - [5.8] $\la$ 0.8 mag  and [3.6] - [4.5] $\ge$ 1.05 mag, zone 3. 

Although the IRAC color-color plot represents a useful tool to search for jets, it has to be used with some caution,
since it does not identify jets uniquely. Indeed, together with the contamination of zone 2 (see above), extinction effects
can be significant in moving points inside and outside the jet {\it locus}. We show 
in Figure~\ref{YL_zones:fig} the extinction vector for A$_V$ = 30 mag; notably, even for extinction values lower than that, Class\,II sources 
may shift inside the jet {\it locus}. Moreover, possible emission of the CO v=1-0 fundamental transition 
at 4.6\,$\mu$m and [FeII] a$^4$F-a$^6$D lines between 4.4\,$\mu$m and 4.9\,$\mu$m can enhance the flux in band 2,
moving the colors toward the top-left corner of the diagram.

Therefore, as also suggested by YL09 themselves, the evaluation of the IRAC colors was complemented by the visual inspection of a three-color map of the region 
(constructed out of the 3.6\,$\mu$m, 4.5\,$\mu$m, and 5.8\,$\mu$m images) to examine the morphology of the emission in the pixels falling in
the jet {\it locus}.

In more detail, this part of the analysis was divided in two steps.
First, the median subtracted IRAC images were converted into a color-map by applying the photometric corrections for extended sources
suggested by the {\it Spitzer} Science Center website for low surface brightness measurements (correction factors of 0.91, 0.94 and 0.71
at 3.6, 4.5 and 5.8 $\mu$m, respectively). As a result, we have obtained a mask for each of the three regions 
previously defined, which were then combined to obtain a false color representation 
(called 'shock-mask') depicted for all the pixels having IRAC colors consistent with shocked H$_2$. 
As an example, 
we show in  Figure\,\ref{IRcolcol:fig}, left panel,
the enlargement of the shock-mask in the region of jet\,\# 4, where the color code is: zone 1: green, 
zone 2: red, zone 3: blue. 


As a second step, we constructed a three-color image by combining the 3.6\,$\mu$m (blue channel), the 4.5\,$\mu$m
(green channel) and the 5.8\,$\mu$m image (red channel), with the aim to confirm (or not) by means of a visual inspection
the possible shock-excited emission localized in the shock-mask. 
First, we discard all pixels that, although falling in the shock-mask, appear in the three-color image as
isolated pixels or structures composed by less than 4-5 pixels which are likely due to artifacts or noise fluctuations.
Then, we applied the following prescriptions to confirm groups of pixels in the shock-mask as real knots:
\begin{itemize}
\item[-] they must appear as extended structures. In SEC10, we define as point-like those sources with a 'shape parameter' $Sh \lesssim 0.7$, where
$Sh$=$\sigma^2 $(observed) - $\sigma^2$(PSF), is the difference between the squares of the half-width of the best-fitting function  and the point-spread function.
Here, we request that knots must have $Sh >$ 0.7. This prescription, although implying that possible point-like knots are discarded, has however the advantage of easily
individuate protostellar sources in the jet {\it locus} of the color-color diagram.
\item[-] preferably, knots should belong to a chain (which can have a straight-, curved-, or $S$-shape). They must be detected at signal-to-noise 
level greater than 3 in at least one band.
\item[-] isolated knots are considered only if detected at signal-to-noise level greater than 3 in at least two bands.
\end{itemize}

In Figure\,\ref{IRcolcol:fig}, right panel, we show the enlargement of the three-color image in the region of jet\,\# 4. This is an illustrative case of the importance
of the visual inspection described above: while the shock-mask in the left panel evidences two possible jets in 
orthogonal directions, the visual inspection of the three-color image confirms the presence only of the jet in the vertical direction, since the pixels in the horizontal direction correspond
to image artifacts, to point-like sources, or to pixels whose photometry is strongly contaminated by the emission of the central cluster.

\subsection{Jet description}
\begin{figure}
\includegraphics[angle=0,width=18cm]{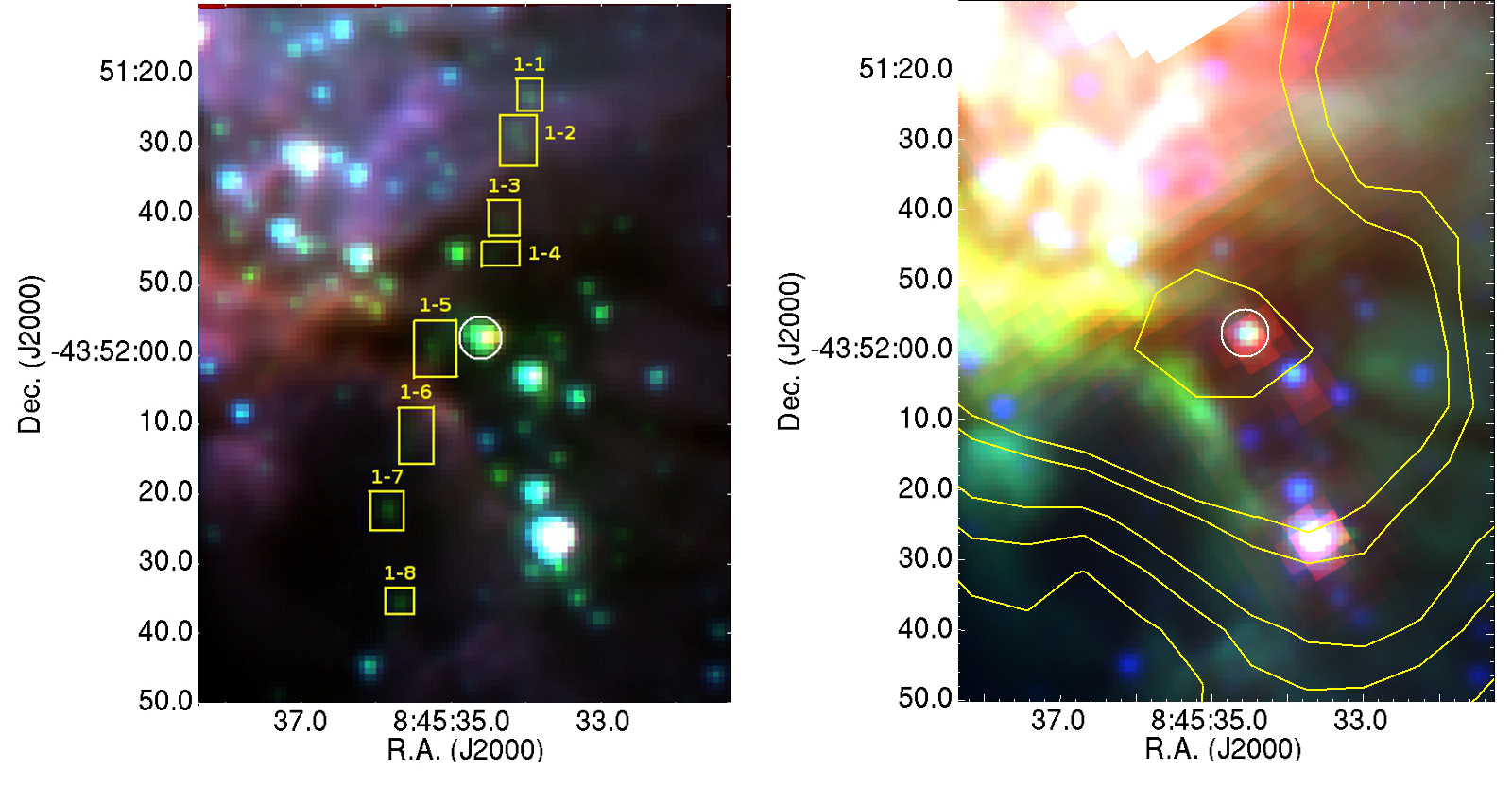}
\caption{Jet 1: {\it Left panel} - Three colors (3.6\,$\mu$m : blue - 4.5\,$\mu$m :
green - 5.8\,$\mu$m : red) IRAC image. H$_2$ knots locations are highlighted with boxes and labeled as in Table\,\ref{tab:tab2}. The 
candidate exciting source(s) is indicated with a circle.
{\it Right panel} - Three colors (3.6\,$\mu$m : blue - 4.5\,$\mu$m : green - 24\,$\mu$m : red)
IRAC/MIPS image where the contours of the dust emission at 1.2 mm (Massi
et al. 2007) are superposed in steps of 4\,$\sigma$ starting
from a level of 50 mJy/beam. The candidate exciting source(s) is (are) indicated with a circle.\label{jet_1:fig}}
\end{figure}

\begin{figure}
\includegraphics[angle=0,width=18cm]{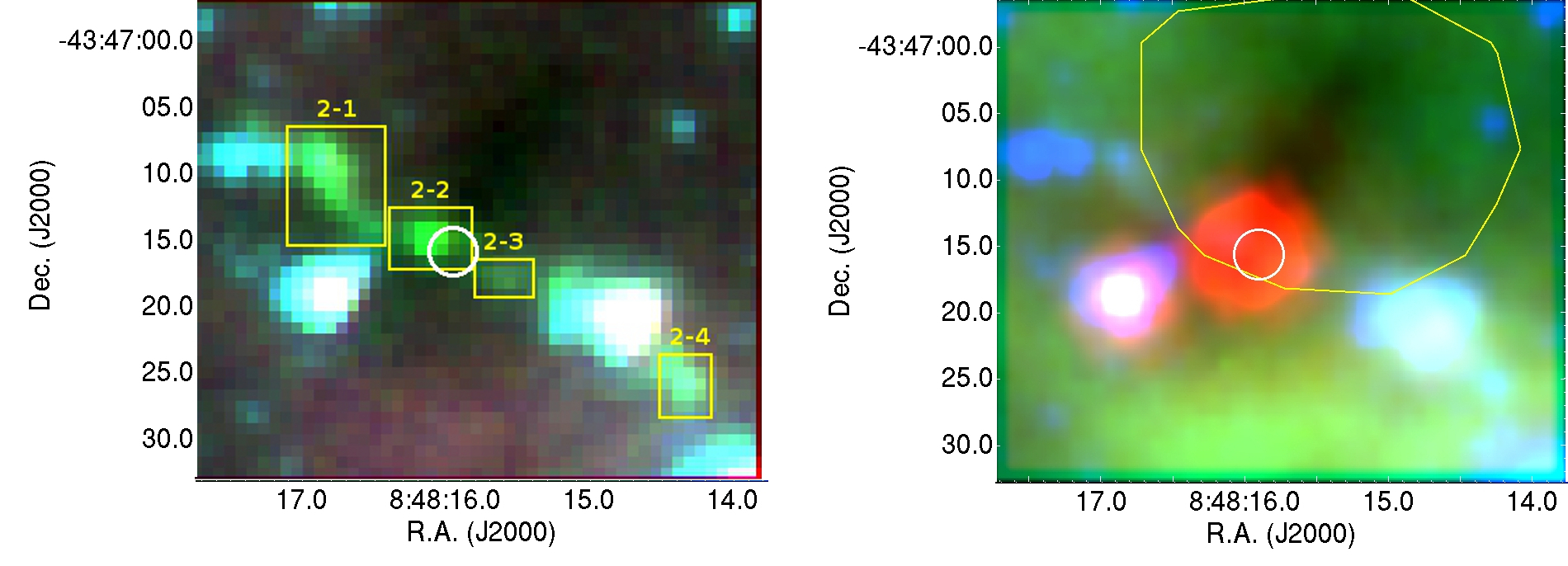}
\caption{As in Figure~\ref{jet_1:fig} for jet 2.\label{jet_2:fig}}
\end{figure}

\begin{figure}
\includegraphics[angle=0,width=18cm]{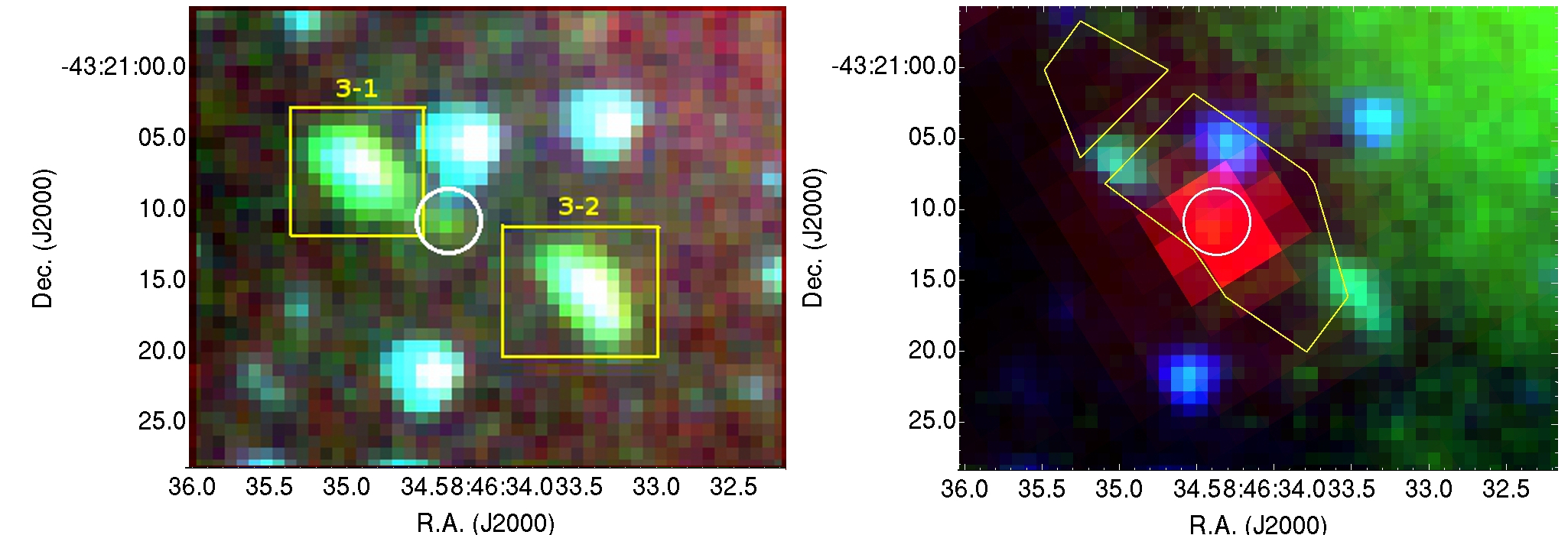}
\caption{As in Figure~\ref{jet_1:fig} for jet 3.\label{jet_3:fig}}
\end{figure}

\begin{figure}
\includegraphics[angle=0,width=18cm]{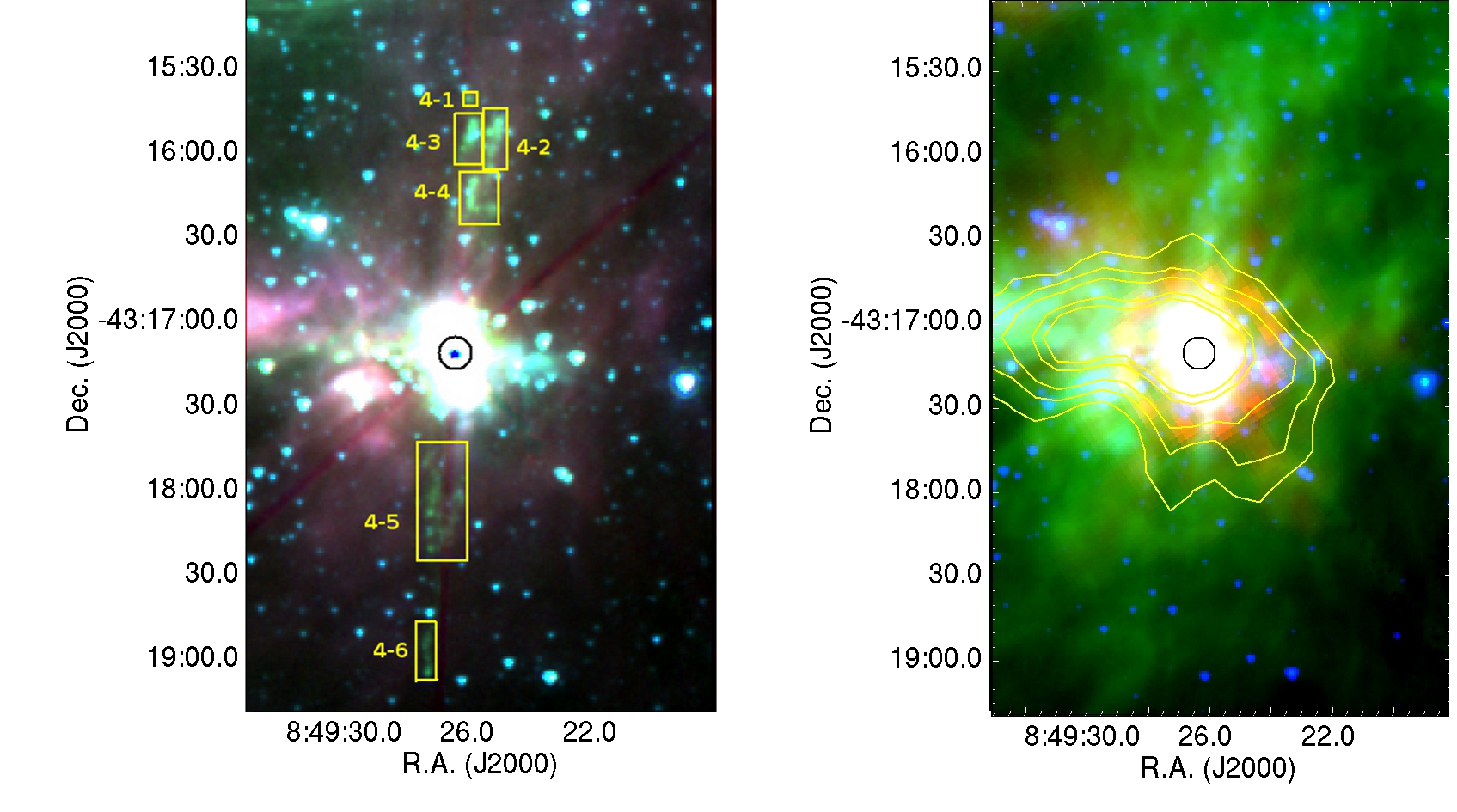}
\caption{As in Figure~\ref{jet_1:fig} for jet 4.\label{jet_4:fig}}
\end{figure}

\begin{figure}
\includegraphics[angle=0,width=18cm]{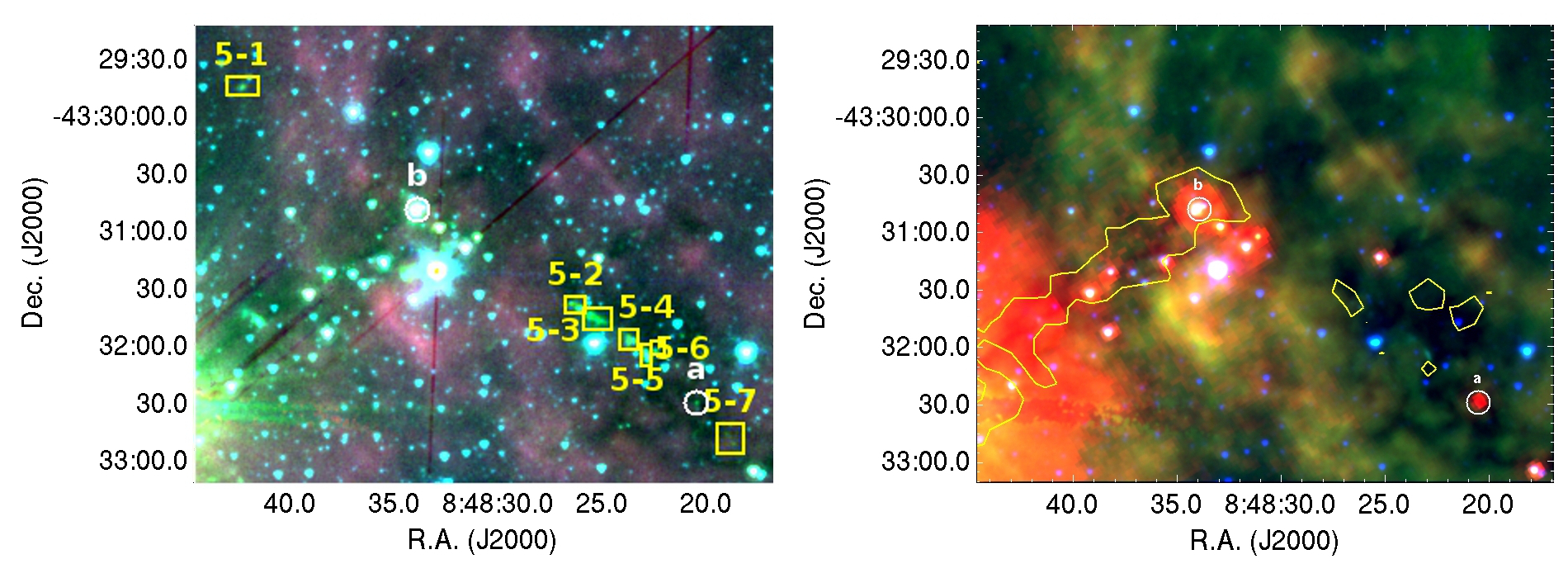}
\caption{As in Figure~\ref{jet_1:fig} for jet 5.\label{jet_5:fig}}
\end{figure}

\begin{figure}
\includegraphics[angle=0,width=18cm]{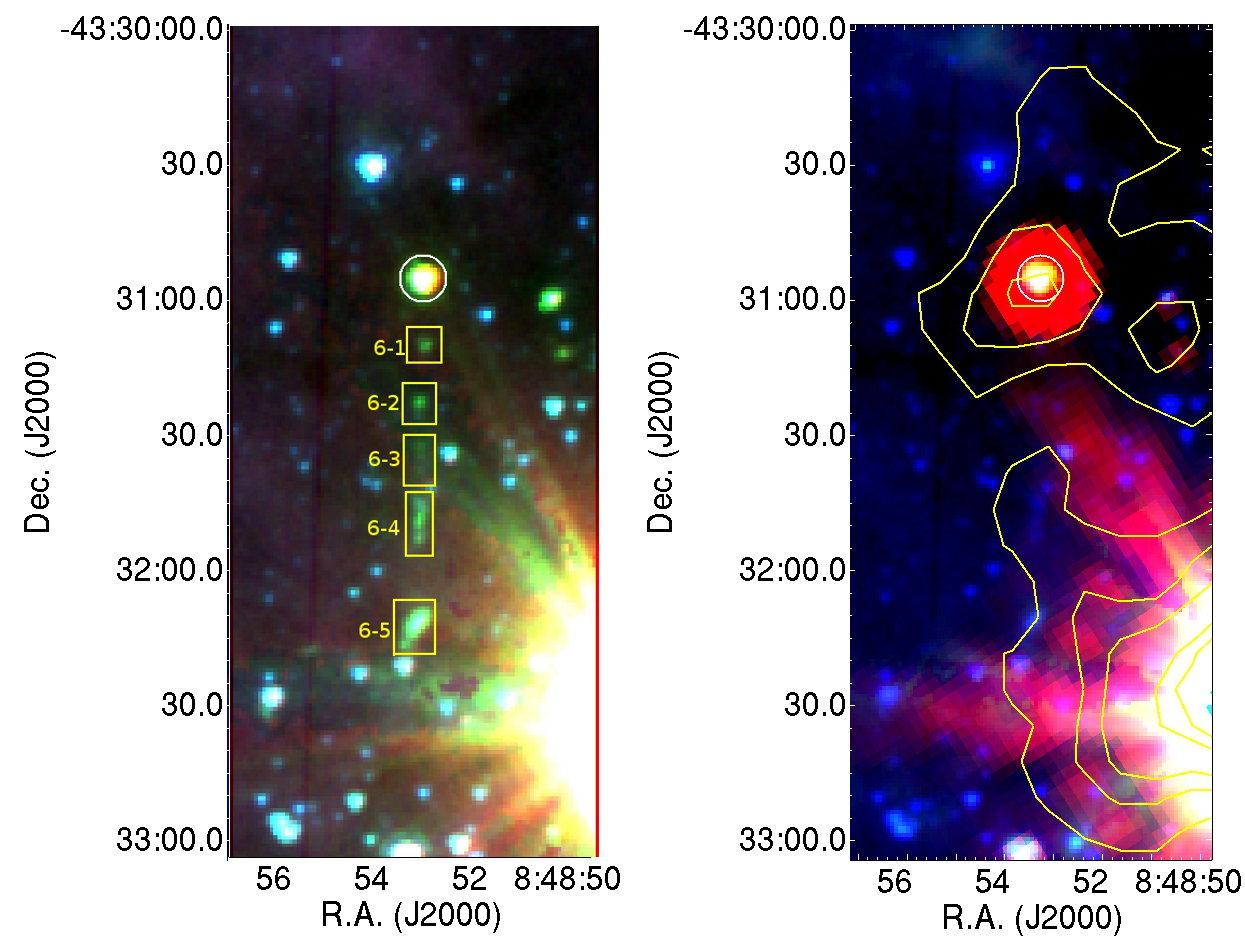}
\caption{As in Figure~\ref{jet_1:fig} for jet 6.\label{jet_6:fig}}
\end{figure}

\begin{figure}
\includegraphics[angle=0,width=18cm]{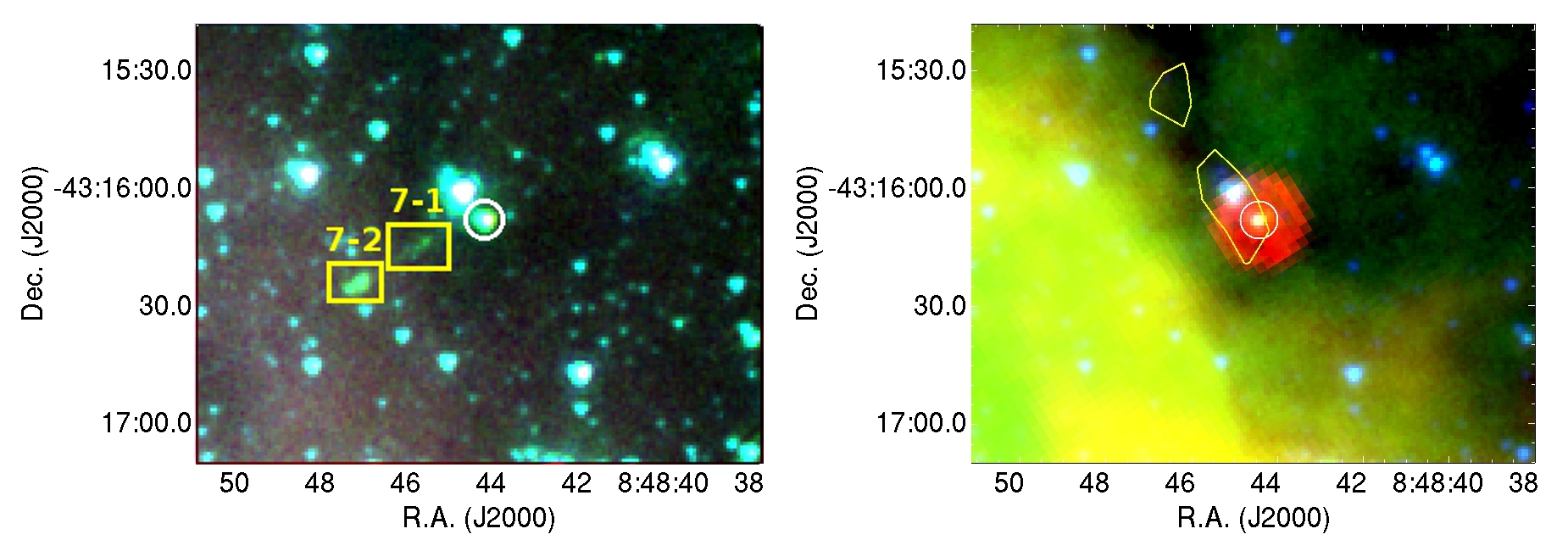}
\caption{As in Figure~\ref{jet_1:fig} for jet 7.\label{jet_7:fig}}
\end{figure}

\begin{figure}
\includegraphics[angle=0,width=18cm]{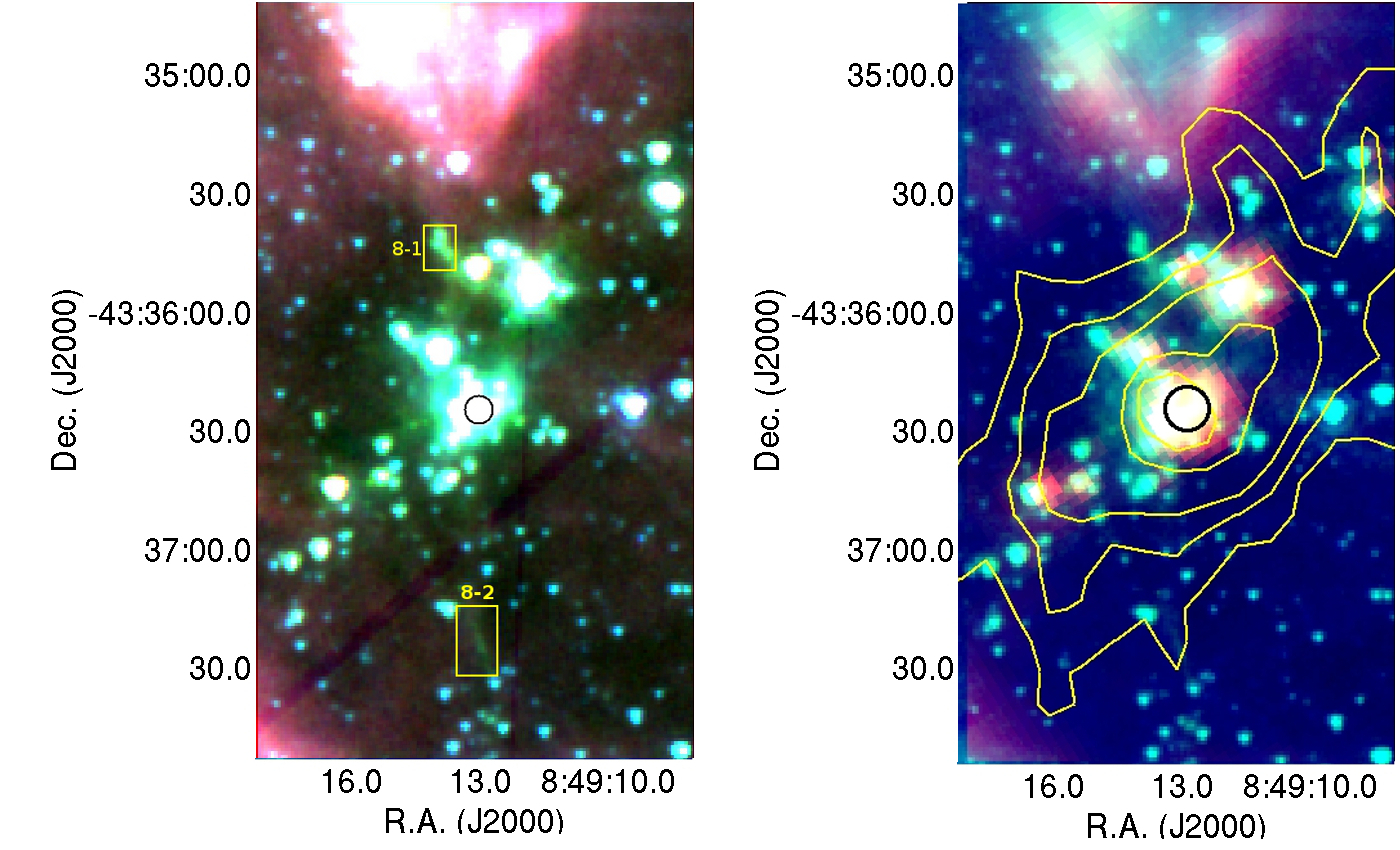}
\caption{As in Figure~\ref{jet_1:fig} for jet 8.\label{jet_8:fig}}
\end{figure}

\begin{figure}
\includegraphics[angle=0,width=18cm]{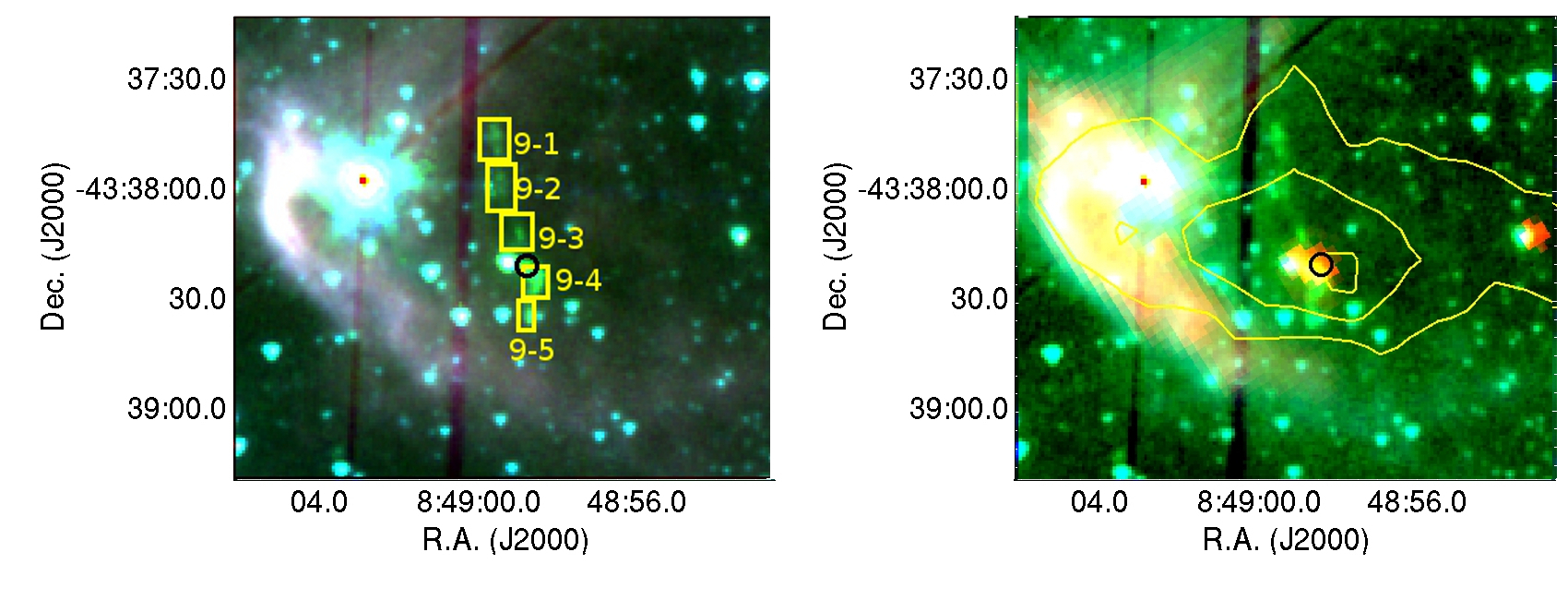}
\caption{As in Figure~\ref{jet_1:fig} for jet 9.\label{jet_9:fig}}
\end{figure}

\begin{figure}
\includegraphics[angle=0,width=18cm]{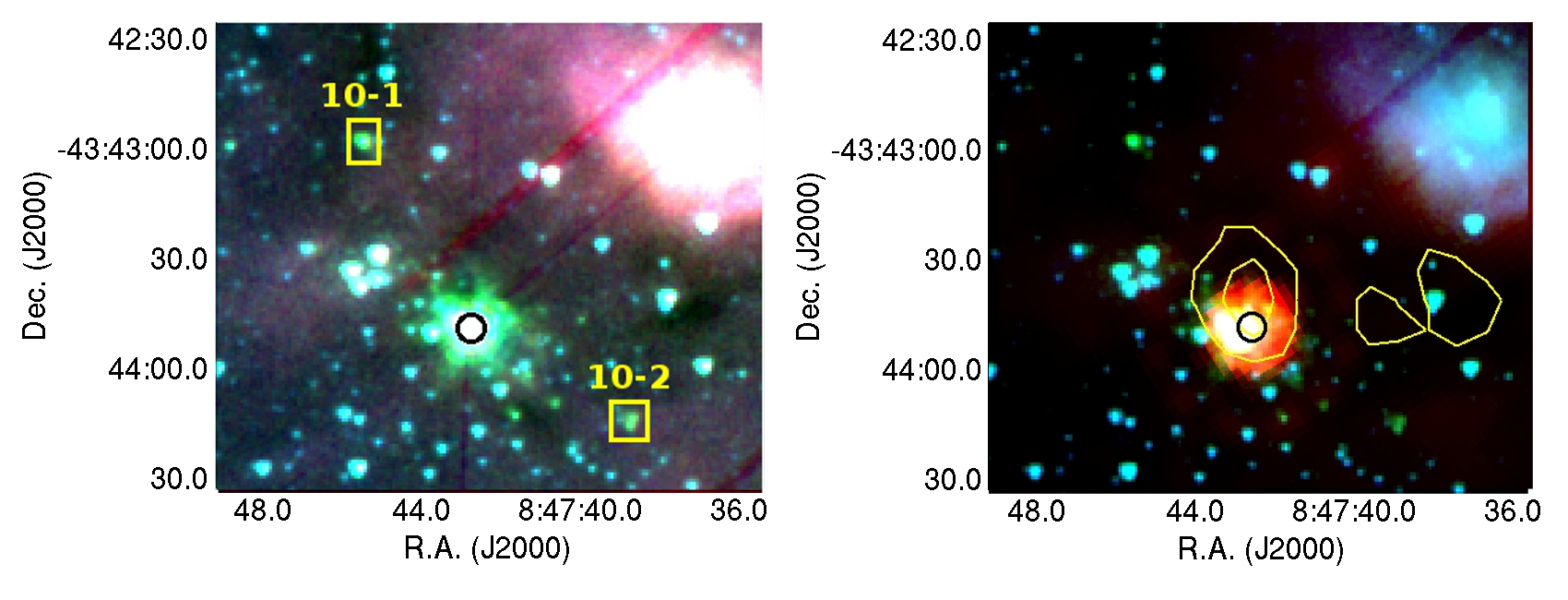}
\caption{As in Figure~\ref{jet_1:fig} for jet 10.\label{jet_10:fig}}
\end{figure}

\begin{figure}
\includegraphics[angle=0,width=18cm]{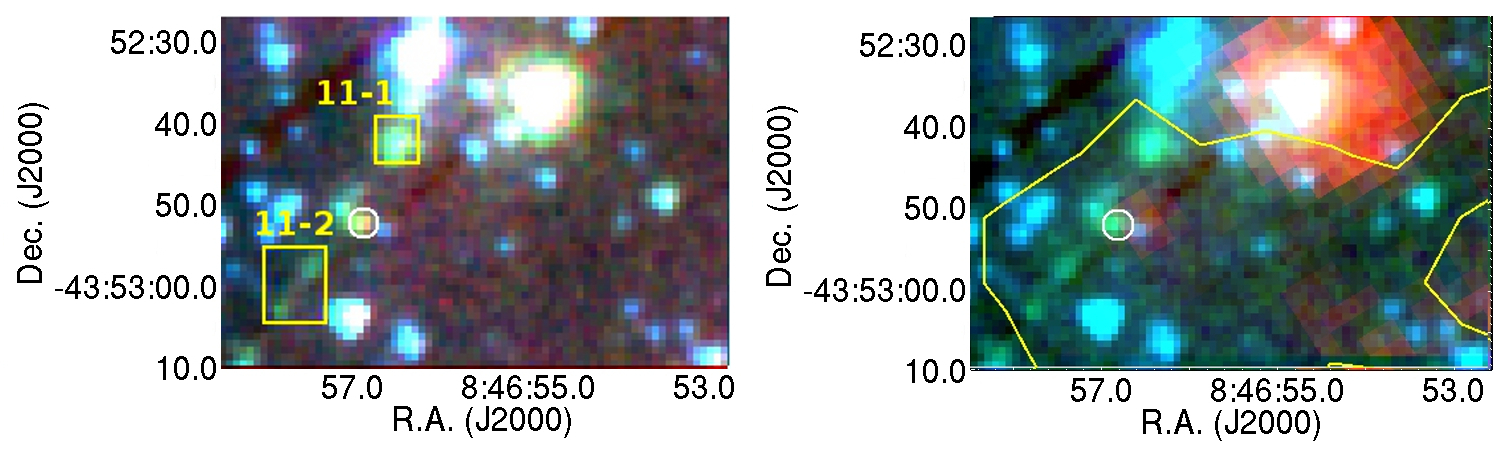}
\caption{As in Figure~\ref{jet_1:fig} for jet 11.\label{jet_11:fig}}
\end{figure}

\begin{figure}
\includegraphics[angle=0,width=18cm]{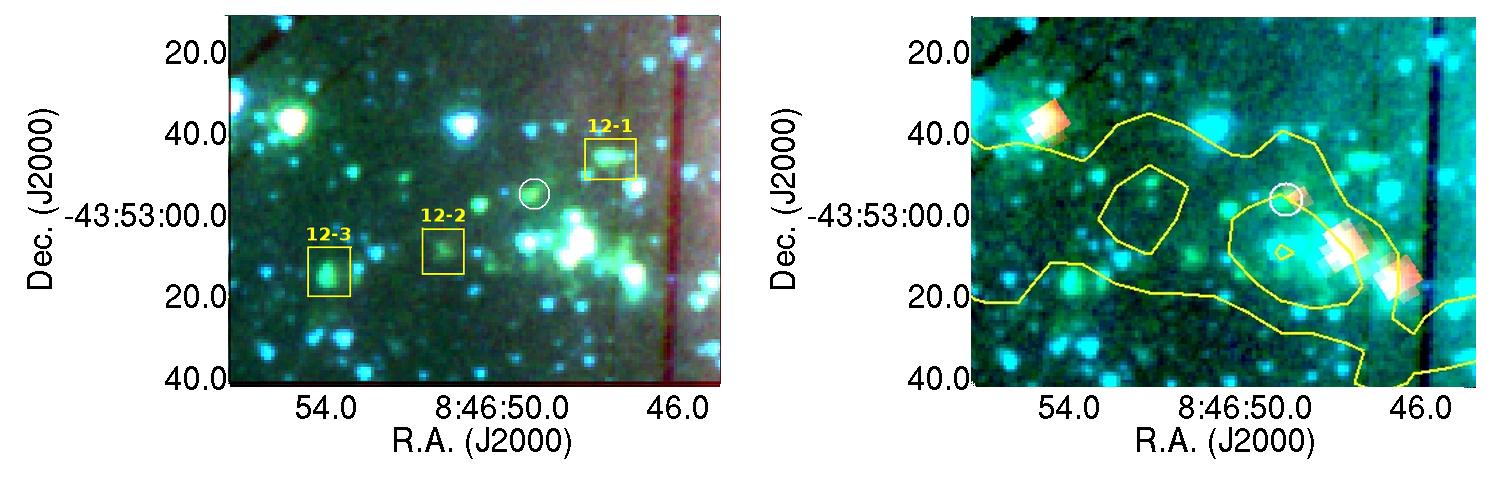}
\caption{As in Figure~\ref{jet_1:fig} for jet 12.\label{jet_12:fig}}
\end{figure}

\begin{figure}
\includegraphics[angle=0,width=18cm]{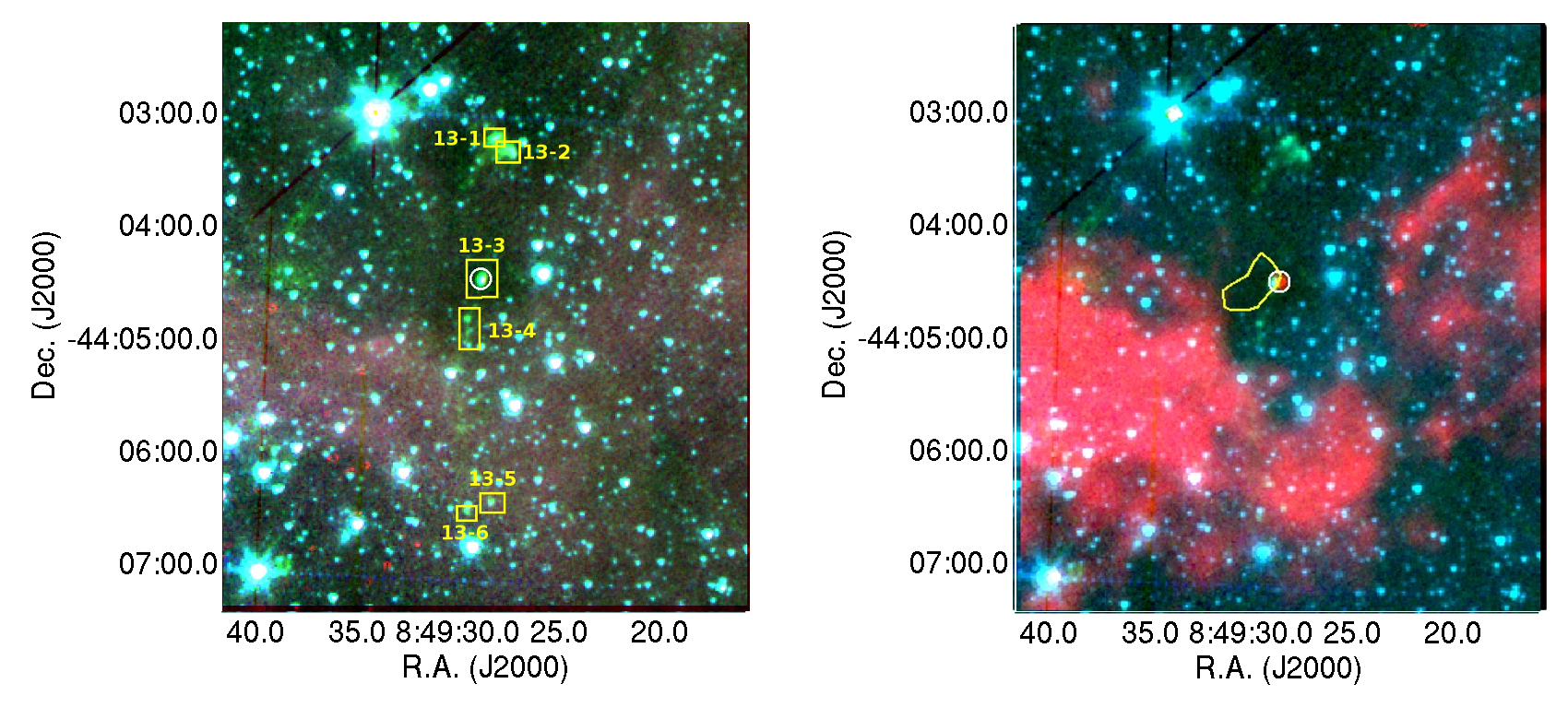}
\caption{As in Figure~\ref{jet_1:fig} for jet 13.\label{jet_13:fig}}
\end{figure}

\begin{figure}
\includegraphics[angle=0,width=18cm]{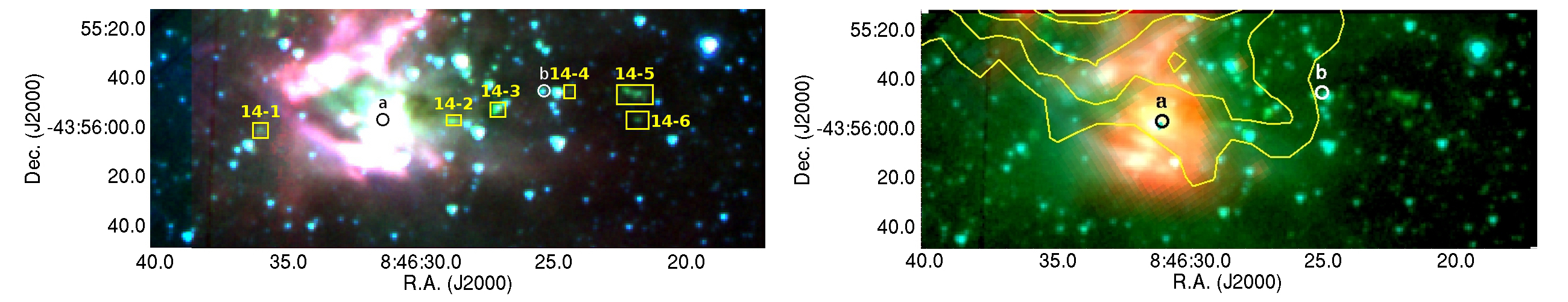}
\caption{As in Figure~\ref{jet_1:fig} for jet 14.\label{jet_14:fig}}
\end{figure}

\begin{figure}
\includegraphics[angle=0,width=18cm]{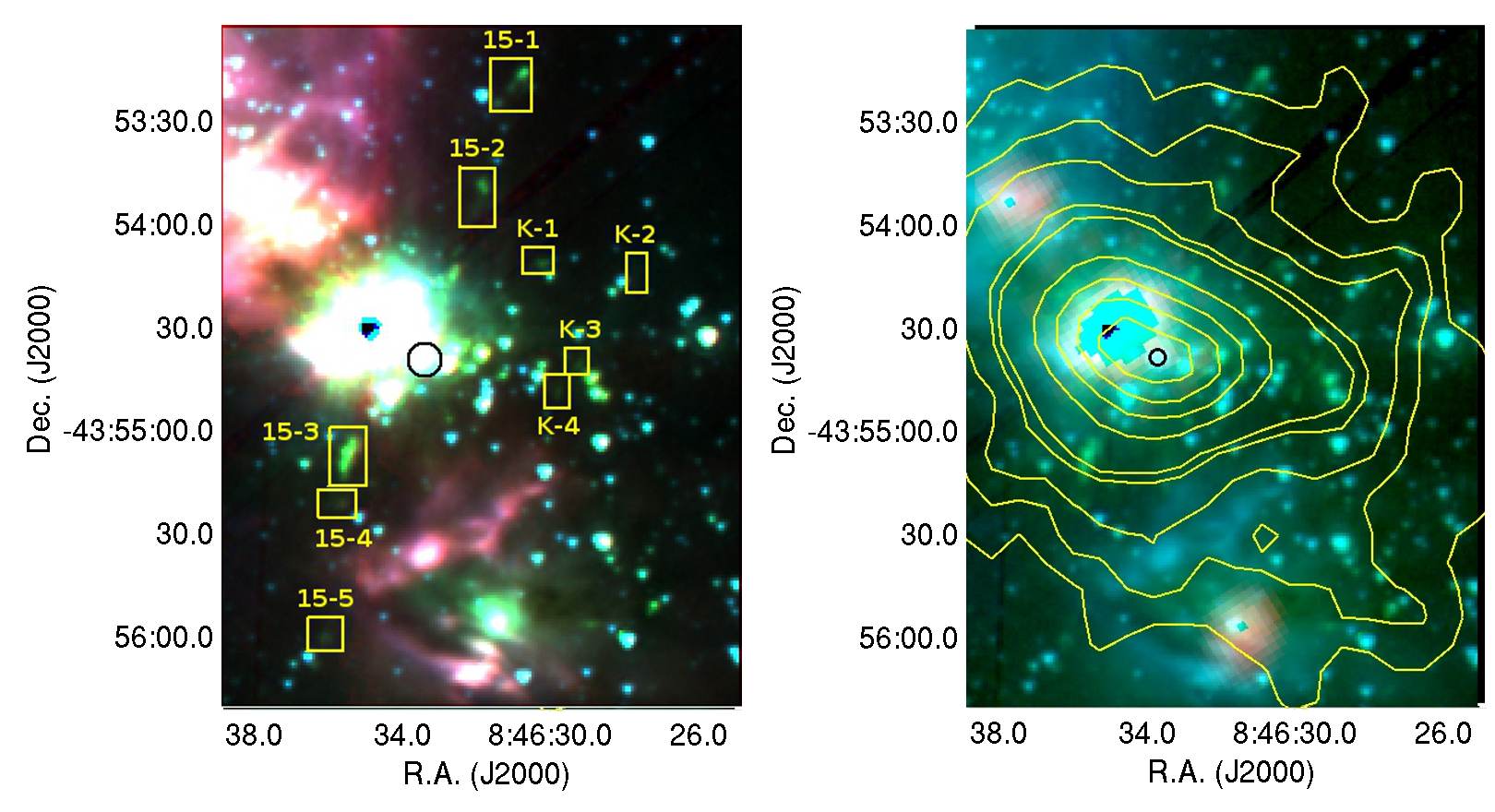}
\caption{As in Figure~\ref{jet_1:fig} for jet 15.\label{jet_15:fig}}
\end{figure}

With the procedure described in the previous section, we have found a total of 15 jets inside 
the area observed with IRAC, as well as some sparse knots that do not show any clear alignment.
Noticeably, we re-confirm with IRAC all the jets that were already detected in the near-infrared, except jets \#\,7 and \#\,14 that represent new discoveries. This is depicted in Figure\,\ref{all:fig}, where it is clear that almost all the \emph{Spitzer} jets (red lines) are located within a field observed in the 
H$_2$ 2.12\,$\mu$m line. Indeed, such result is not unexpected, since H$_2$ \emph{Spitzer} and 2.12\,$\mu$m lines, coming from levels with similar excitation energy, probe 
similar excitation conditions.
Conversely, some knots (within the jets) detected in the 1-0 S(1) line with SofI do not 
appear in any \emph{Spitzer} image. In the case of jet \# 15, this occurs because the most internal knots (e.g. knots D and F reported by
Giannini et al. 2005) are confused in the \emph{Spitzer} images in the strong diffuse emission of the protostellar cluster where the
exciting source is located. 
In other cases of faint knots (at 2.12\,$\mu$m), the lack of a \emph{Spitzer} counterpart can be ascribed to the poorer sensitivity at
longer wavelengths. For example, the 3\,$\sigma$ sensitivity limit at 2.12\,$\mu$m of $\sim$ 0.01 mJy is around a factor of 3 better than the 
sensitivities at 3.6\,$\mu$m and 4.5\,$\mu$m.  
 
The very nice coincidence between IRAC and SofI jet detections leads to three main conclusions: {\it i}) 
the approach adopted to discover jets in the IRAC images can be considered as validated, {\it ii}) H$_2$ 2.12\,$\mu$m surveys allow to 
catch a relevant fraction of jets and hence longer wavelengths observations are not mandatory to overcome extinction problems, and, {\it iii}) the absence of
jets in the parts of the IRAC map not covered also in the near-infrared strongly indicates that jets can be preferentially found
in proximity of the dust peaks, namely where the near-infrared images were originally taken (see also De Luca et al. 2007).  

The 15 \emph{Spitzer} jets are depicted in Figures\,\ref{jet_1:fig} to \ref{jet_15:fig} and are described in Appendix \ref{sec:appendix}. 
In each of these figures, we show in the left panel a composite IRAC 3-color image and
in the right panel a composite IRAC-MIPS image with overlaid the contours at 
1.2 mm (these latter in steps of  4\,$\sigma$ starting from a 4\,$\sigma$ level of 50 mJy/beam). In these figures, jet knots are marked by small boxes while the candidate exciting source, determined as described in Sect.\,\ref{sec:sec4}, is marked by a small circle.

A summary of the jet properties is given in Table~\ref{tab:tab1}, where for each jet we give in columns 1-4 the identification 
number\footnote{The jets identification number follows our internal classification.}, the number of the detected knots, the projected 
length, and the position angle. 
As anticipated, all the jets were originally discovered in the H$_2$ 2.12\,$\mu$m line, the only exceptions being represented by jet\,\#7,
\#\,14 and by other few knots (K1-K4), which were not targeted in the near-infrared. In the last column, we give the references where these jets have been reported. 
The projected jet lengths and position angles are in agreement with those given in the literature, the only exception being the length of 
jet\,\#\,6, which was originally supposed to have a double length with respect to that reported in Table\,\ref{tab:tab1} (Giannini et al. 2007). This hypothesis was done since the jet appears 
one-sided in the 2.12\,$\mu$m image, and the lack of a counter-jet was ascribed to the limited field of view of SofI. Based on the IRAC image (see Fig.\,\ref{jet_6:fig})
we now confirm the lack of any additional knot symmetric to the candidate ES.    
As a general note, jets \# 2\,,3\,,7, and 11 are rather compact (length lower than 0.15 pc), jets \# 4\,\,,14, and 15 extend for more than 0.4-0.5 pc, while all the others have intermediate lengths.
Length of jets \#5 and \# 13 is very uncertain, since it is not clear whether or not the discovered knots can be effectively attributable to the same jet (see also the discussion in Appendix\,\ref{sec:appendix}).
Regarding the position angle (PA, computed from the North to the East), notably around half of the jets are oriented 160$^{\circ} \pm 15^{\circ}$ : we will comment this feature in Sect.\,\ref{sec:sec5.4}.

\subsection{Jet photometry}
To obtain the photometry of the detected knots we have applied the task {\it polyphot} within the Image Reduction and Analysis Facility (IRAF) package,\footnote{Available at http://iraf.noao.edu/}  which is suited 
for evaluating the emission of extended and irregular sources. For each knot (both in the IRAC bands and in the narrow-band at 2.12 $\mu$m),
we have first defined a polygon with sides tangent to the 3\,$\sigma$ contour of the emission and then summed up the flux of all the pixels inside it. The 
sky background was estimated locally and then subtracted from the flux computed inside the polygon. As shown by Neufeld \& Yuan (2008), 
the photometry in the IRAC bands are dominated by bright H$_2$ lines, and in particular by the $v$=1-0 O(5)-O(7) in band 1 
(3.6\,$\mu$m), $v$=0-0 S(9) in band 2 (4.5\,$\mu$m), $v$=0-0 S(7)-S(6) in band 3 (5.8\,$\mu$m), and $v$=0-0 S(5)-S(4) in band 4 (8.0\,$\mu$m).
Minor contributions are expected from PAH, CO and atomic lines which, however, cannot be easily estimated (Takami et al. 2010) and thus are not taken into account by us in the
photometry. Similarly, we ascribe the near-infrared narrow-band filter photometry totally to
H$_2$ 1-0 S(1) line emission, since no continuum emission in the $K$-band is usually seen in the spectra of protostellar jets (e.g. Giannini et al. 2004, Nisini et al. 2002). 
Table\,\ref{tab:tab2} lists the coordinates and the size of each knot, evaluated on the 4.5\,$\mu$m image, along with the photometry between 2.12\,$\mu$m and 8.0\,$\mu$m.
The last two columns list the IRAC luminosity ($L_{\rm{IRAC}}$, obtained by summing up the luminosities of all the IRAC bands) and 2.12\,$\mu$m luminosity ($L_{1-0S(1)}$), both computed assuming a distance to VMR-D of 700 pc (Liseau et al. 1992).


\section{Exciting sources analysis}\label{sec:sec4}
\subsection{Exciting sources identification}
As a second goal of our study, we have used the IRAC and MIPS images to search for the ES using
the following procedure.

\begin{itemize}
\item[-] On the IRAC image at 4.5\,$\mu$m we have defined a rectangle oriented as the jet
itself and having the long side equal to $n$ $\times$ the jet length, where the value of $n$, in the range 1-3, is 
selected for each jet to not exceed the parsec scale. The short side, orthogonal
to the jet direction, is defined case-by-case so that all the knots lie inside the rectangle. 

\item[-] We then searched the IRAC catalog (SEC10) for all the point-like sources inside the rectangle,
selecting those whose position is compatible with the jet shape (straight- curved- or {\it S}-shape). 

\item[-] Sources are then selected if they have a spectral index (computed by using the available measurements between $K_s$ and 8\,$\mu$m) 
compatible (within the upper limits) with a YSO of Class\,II, I or flat. To Class I or flat sources we give higher priority than to nearby
Class II sources.
Analogously, we have searched the MIPS catalog to select possible sources not present in the
IRAC catalog because not detected at wavelengths shorter than 24\,$\mu$m.  

\end{itemize}

The results of this search are presented in Table\,\ref{tab:tab3}. We remark that, given the crowding of YSOs in the investigated fields, 
we can not consider our search exhaustive, especially for the presence of many Class\,II sources in the defined rectangles. Consequently, the sources
listed in Table\,\ref{tab:tab3} have to be considered as ES {\it candidates} and hence as interesting objects to be further studied.
Noticeably, however, we find only one best candidate in all cases, apart from jets \#\,5 and \#\,14, for which two possible candidates have been selected 
(labeled with {\it a} and {\it b} according with a priority order, see Appendix\,\ref{sec:appendix}).  

Associations with counterparts at different wavelengths (from 1.2\,$\mu$m to 1.2 mm) are reported, with the
corresponding identification number assigned in each catalog. Associations between IRAC, MIPS and SIMBA sources
have been already done by SEC10 and reported in their photometric catalog; here we also consider
possible associations with 2MASS, WISE and BLAST sources. These have been matched to the IRAC source list by using a circle whose radius is the 
sum of the positional uncertainties associated with each pair of instruments (reported in the caption of Table\,\ref{tab:tab3}).
The ES of jets \#\,2, \#\,3, \#\,11, \#\,13 are very barely or not detected at wavelengths shorter than 24\,$\mu$m;
this confirms the results presented in Giannini et al. 2007 (for jets  \#\,2, \#\,3) who hypothesized them as very young YSOs.  

All except one sources have a MIPS 24\,$\mu$m counterpart and 8 (10) are associated with a 250\,$\mu$m 
(1.2 mm) peak, while only 5 objects are seen in the 2MASS bands. The source exciting jet\,\#\,15 has been already the subject of a dedicated study (Giannini et al. 2005). It is a member of an infrared cluster that appears completely unresolved and saturated in the
\emph{Spitzer} bands. For this reason that object (also known as IRS\,17-\# 40-3) will not be commented further on in the present paper.

In Table\,\ref{tab:tab4} we list the fluxes (in mJy) of the ES, which will be used in Sect.\,\ref{sec:seds} to derive their SED. As far as the WISE data are concerned,
we give only the fluxes at 12\,$\mu$m (band 3). Those in bands 1,2,4 (at 3.4 $\mu$m, 4.6 $\mu$m, and 22 $\mu$m) do not differ from the \emph{Spitzer}
ones (at 3.6\,$\mu$m, 4.5\,$\mu$m, and 24\,$\mu$m) by more than 30\%, which does not
imply significant modifications in the SED global shape. Two sources (\#\,4 and \#\,14a) are saturated at 24\,$\mu$m, hence the WISE\,4 band flux has 
been used for their SED fitting. Finally, in the last column of Table\,\ref{tab:tab4}
we list the source spectral index, computed for sources with at least four valid fluxes between 2.16\,$\mu$m and 12\,$\mu$m.

\subsection{Evaluation of the jet extinction}\label{sec:jetext}

As a further step toward identifying the driving sources, we have investigated whether the selected candidate ES of each jet is located at the center of two distinct lobes traveling in opposite 
directions, as expected for bipolar jets. Since we lack high-resolution spectroscopy observations, which would have allowed us to directly measure the mutual shift of the emission lines, we have adopted
an indirect way to identify (if any) the blue- and red-shifted lobes. This method is based on the comparison of the IRAC total luminosity ($L_{\rm{IRAC}}$) with the luminosity of the
1-0 S(1) line ($L_{1-0S(1)}$), both listed in the last two columns of Table\,\ref{tab:tab2}. Such comparison helps in identifying 
the blue- and red-shifted lobes under some conditions: {\it i)} the visual extinction increases in the direction of the red lobe (as expected 
if this coincides with the receding part of the jet and if this latter is not in the sky-plane); {\it ii)} $L_{1-0S(1)}$/$L_{\rm{IRAC}}$ weakly depends on the temperature and density in a given knot.\\ 

To verify the condition {\it ii)}, we have estimated the ratio $L_{1-0S(1)}$/$L_{\rm{IRAC}}$ for temperatures between 1000 K and 3000 K, in the Local Thermal Equilibrium (LTE) approximation. As shown in Figure\,\ref{LTE:fig}, such ratio remains fairly constant in the range $\sim$ 0.2-0.4 (note that this is generally not true if the ratio of $L_{1-0S(1)}$ with the luminosity in a single IRAC band is taken). Hence, a difference of  $L_{1-0S(1)}$/$L_{\rm{IRAC}}$ of more than a factor of $\sim$ 2 in symmetrically located knots
(with respect to the putative exciting source) can be primarily ascribed to a difference in the local extinction ($\Delta$A$_V$), since this latter affects the near-infrared and the mid-infrared fluxes in a differential way. (We note, however, that the assumption {\it ii)} is too coarse to allow the absolute determination of A$_V$).\\

In Table\,\ref{tab:tab5} we list for each jet the pairs of symmetric (groups of) knots for which we have evaluated 
$\Delta$A$_V$, computed by assuming the Rieke \& Lebofsky (1985) extinction law. To minimize the uncertainties, this has been obtained from  the average $L_{1-0S(1)}$/$L_{\rm{IRAC}}$ in each side of the jet. 
Adding the photometric uncertainties 
to the allowed variations in $L_{1-0S(1)}$/$L_{\rm{IRAC}}$ due to the physical conditions in the knots, we consider as significant only $\Delta$A$_V$ larger than 2 mag, which roughly correspond to a difference of $L_{1-0S(1)}$/$L_{\rm{IRAC}}$ of more than 4.
Summarizing the results, our objects are divided in three groups, in which we measure a $\Delta$A$_V$ with a decreasing level of significance.
In particular we get:
 $\Delta$A$_V$ $>$5 mag in five jets (\# 1, 2, 4, 8, 11), 2 mag $<$ $\Delta$A$_V$ $\le$ 5 mag in two jets (\# 3, 12) and  $\Delta$A$_V$ $\le$ 2 mag in four jets (\# 5, 9, 10, 15). For jet \#\,13 we measure $\Delta$A$_V$ both considering knots 3 and 4 ($\Delta$A$_V$= 10 mag) and the groups of knots 1-2 and 5-6 ($\Delta$A$_V$= 4 mag). In both cases all the knots 
south (north) the ES are red- and blue-shifted, respectively. This result slightly favors the hypothesis that all the knots are part of the same jet (see also Appendix\,\ref{sec:appendix}).
For the remaining jets we are not able to apply the method either because we can not identify knots symmetrically located (jet \# 6) or because measurements of the 2.12\,$\mu$m line
do not exist (jets \# 7, 14). In particular, for jet \# 14 (as well as for jet \# 5, where $\Delta$A$_V$ is not significant) we can not use the 'extinction method' to identify the exciting source 
between the two candidates labeled with $a$ and $b$ in Table\,\ref{tab:tab3}. 

Remarkably, the few direct velocity measurements existing in the literature for the VMR-D jets are all in agreement with the results obtained with the above 'extinction method'. These are the high-resolution spectroscopy observations of the 2.12\,$\mu$m line of jet\,\# 15 (Giannini et al. 2005), the $^{12}$CO(1-0), $^{13}$CO(2-1) maps in the vicinity of the IRAS sources and overlapping jets \# 4, 8, and 15  (Elia et al. 2007), and the small $^{12}$CO(3-2) maps obtained with APEX for jets \# 1, 2, and 3 (see as an example the map of jet\,\# 3 shown in Figure\,\ref{fig:apex}). Another $^{12}$CO(3-2) map (although incomplete) of jet \# 5 should favor source 5b as the exciting source, since all the 
parts of the outflow overlapping knots 2-7 appear blue-shifted.


\section{Discussion}
\subsection{SEDs fits}\label{sec:seds}

Once the observed fluxes have been determined to construct the SEDs for our candidate ES, 
we used the radiative transfer model developed by Whitney et al. 2003a, Whitney et al. 2003b 
to obtain insight on the physical properties of these objects. 
Actually, the use of this model has been greatly facilitated by Robitaille et al. 2006, Robitaille et al. 2007, 
who provided a grid of 200,000 computed models and an efficient online fitting procedure\footnote{Available at http://caravan.astro.wisc.edu/protostars/sedfitter.php} to help the community in the SED fitting work. 
In the best-fit procedure we set as constraints the extinction between
0 mag and 200 mag and the distance to the VMR-D within the range 500 pc - 1000 pc (Liseau et al. 1992). 
We fit the SED of sources with at least six photometric detections in Table\,\ref{tab:tab4}. In addition, a single grey-body function was used to fit the SED of the sources 
not observed at wavelengths shorter than 24\,$\mu$m,  but with at least 3 data points (\# 2, \# 3, \# 13).
We used different-sized apertures with radius corresponding to 3\,$\sigma$ of the instrument beam and verified that even adopting larger apertures the same models are selected. 
The results of the best-fits 'a la Robitaille' are depicted in Figure\,\ref{seds:fig}. Here we show as upper limits also the BLAST (350-500\,$\mu$m) data points, although not considered in the fitting procedure. This choice is motivated by the fact that multiple sources lie in the large BLAST beam. The main output parameters are reported in Table\,\ref{tab:tab6}, where we give both the values of the best-fit model (in boldface) and those of the model whose $\chi^2$ is 50\% higher than the minimum value. These parameters are the distance to the source (col.2), the disk inclination (between 0$^\circ$ and 90$^\circ$ with respect to the sky-plane, col.3), the interstellar and circumstellar extinction (cols.4,\,5), the stellar temperature and mass (cols.6,\,7), the outer envelope radius (col.8) along with its mass (col.9), and the total (stellar plus accretion) luminosity (col.10). The general picture emerging from the SEDs modeling is that most of the
ES are low-mass and relatively evolved protostars.  The stellar temperatures (2900 K $\la T_\star \la$ 4400 K)  and masses (typically $M \approx$ 0.05-1 M$_\sun$) are those typical of embedded (A$_V$ $\sim$ 10 - 60 mag) T Tauri stars of spectral type between K0 and M0. The reservoir of material of the surrounding envelope ($M_{\rm{env}}$ typically between 10$^{-2}$- a few M$_\sun$) is not enough to enhance significantly the final stellar mass. Possible exceptions to this general picture are : {\it i)} sources \# 1 and \# 4, whose mass (already piled up onto the star or still in the envelope) appears potentially sufficient to form an intermediate-mass object. Given the crowding of the regions where these sources are located (close to infrared clusters, see also Table\,\ref{tab:tab1}), we can not exclude however that they are binaries or multiple objects unresolved at the IRAC angular resolution; {\it ii)} sources \# 7 and \# 14a, for which the fitted  $M_{\rm{env}}$ and $R_{\rm{env}}$ are very uncertain, since their SED is not measured beyond $\lambda$=70\,$\mu$m.
\textsf{•}Finally, we note that the disk inclinations of sources \# 1, \# 4, \# 8,
for which the correspondent jets have two mutually shifted lobes (i.e. for which $\Delta$A$_V$ is $>$ 5 mag, see Sect.\ref{sec:jetext}), are 20$^\circ$-40$^\circ$, 55$^\circ$-65$^\circ$, 20$^\circ$. In the hypothesis that jet and disk are orthogonal, these inclinations confirm that these jets lie outside the sky-plane. 

\subsection{Jets vs. exciting sources properties}\label{sec:sec5.3}
In Figure\,\ref{fig:jetcolor} the projected jet length (in pc) is displayed as a function of the far-IR color [24]-[70]\footnote{The MIPS [24]-[70] color is defined as 2.5log($F_{70}/F_{24}\times F0_{24}/F0_{70}$) where $F_{70}$,\,$F_{24}$ are the fluxes at 24, 70 $\mu$m, respectively and $F0_{70}$,\,$F0_{24}$ the zero magnitude fluxes.} 
of the candidate ES.
This plot is based only on sources selected as first priority ES (see also Appendix\,\ref{sec:appendix}). The idea underlying such plot is to 
investigate whether or not a relationship exists between the two considered parameters within 
an evolutionary framework. Indeed, the youngest jets
are expected to be the shortest, in the assumptions that all the jets travel at comparable velocities and that their inclinations 
with respect to the sky-plane are not dramatically different. In this sense the following analysis has to be considered in a statistical sense, and no conclusions 
can be drawn for the individual sources.
The [24]-[70] color is typically used as an age indicator, because the youngest sources are also the coldest 
and reddest ones. Figure \ref{fig:jetcolor} shows an evident correlation (regression coefficient $>$ 0.8) which well agrees with the 
expectations. Noticeably, the same kind of plot constructed with IRAC (instead of MIPS) colors or with the spectral index 
between 2\,$\mu$m and 12\,$\mu$m
does not show any correlation: bluer colors are, in fact, more sensitive both to different circumstellar 
morphologies (e.g. disk inclination) and to extinction effects. On the other hand, the [24]-[70] color is more sensitive to the circumstellar 
envelope temperature and almost unaffected by extinction, 
since it changes of only 0.2 mag for A$_V$ $\approx$ 100 mag. Hence, the observed correlation is 
compatible with an evolutionary interpretation, and can not be related to a mere inclination effect: if this latter 
were responsible for the shortening of the jet projected length, the 
bluest (warmest) portions of the circumstellar disk material would be exposed to the observer, hence an anti-correlation would emerge
between the jet projected length and the [24]-[70] color.

In Figure\,\ref{fig:lumlum} we plot the IRAC line cooling as a function of the ES bolometric luminosity (considering only ES of higher priority). Our aim is to investigate the balance 
between the energy irradiated by the driving  protostar and that emitted by the associated jet. In that Figure, the horizontal dotted lines are the range of bolometric luminosity given in Table\,\ref{tab:tab6}, with the value associated with the best-fit model indicated with a dot. The reported values of $L_{\rm{IRAC}}$ are those given in Table\,\ref{tab:tab2} without any extinction correction; for reference, these values increase by a factor of 0.3 and 1.3 (in logaritmic scale) for A$_V$ = 10 mag and 30 mag, respectively. Instead, the uncertainties on $L_{\rm{IRAC}}$ associated with the photometric errors are not significant.
A rough linear correlation (in the log-log scale)
is found between the two quantities (regression coefficient of 0.68), with an intercept value located at $L_{\rm{IRAC}}$ $\sim$ 10$^{-1.96}$ L$_{\sun}$). This value can be easily converted in the total H$_2$ cooling ($L_{\rm{H_2}}$) by estimating the ratio $L_{\rm{H_2}}/L_{\rm{IRAC}}$ under LTE conditions: we find that this ratio is $\sim$ 4 at temperatures of 2000-3000 Kelvin.
Hence, the above relation translates into $L_{\rm{H_2}}$ $\sim$ 10$^{-1.56}$ L$_{\sun}$.
This value is in good agreement with the result ($L_{\rm{H_2}}$ $\sim$ 10$^{-1.4}$ L$_{\sun}$) found by Caratti o Garatti et al. (2006) in a sample of Class\,0/I sources. 
In that work no significant differences in terms of efficiency were found between the two classes, likely because of the large uncertainties associated with $L_{\rm{H_2}}$, which was
largely derived on the base of near-infrared observations. These differences are indeed expected in the commonly accepted scenario in which a tight relationship exists between mass-loss and mass accretion rate, with a declining of the latter with the source evolution (e.g. Bontemps et al. 1996),
In Figure\,\ref{fig:lumlum} a certain degree of segregation between the points above (sources \# 1, 3, 4, 6, 9, and 13) and 
below the straight line (sources \# 2, 7, 8, 10, and 14a) is indeed recognizable, the formal fits through the two groups having regression coefficients of 0.97 and 0.99, respectively. We estimate that the two groups can mix each other only if their differential extinction is (on average) larger than 15-20 mag. 
Even a more clear segregation is found by using far-infrared lines. For example, the ratio between 
$L_{\rm{FIR}}$ (i.e. the cooling rate due to the far-infrared lines) with  $L_{\rm{bol}}$ gives values spanning from 10$^{-2}$ to 10$^{-4}$ going from Class\,0 to Class\,II sources (Nisini et al. 2002, Lorenzetti et al. 2002a).

\subsection{Large-Scale Properties}\label{sec:sec5.4}
The circumstellar accretion disks are expected to have their
symmetry axis parallel to the local magnetic field (Shu et al.
1995). 
At larger scales, the magnetic field is expected to have a role only
in regions located in the Galactic plane and far from the center, i.e. where   
it is oriented parallel to the Galactic disk (150$^{\circ}$-160$^{\circ}$, Stephens et al. 2011).
Indeed, jets with orientations parallel to the large-scale magnetic field
have been already found both {\it in} the Galactic plane (Hodapp
1984, 1987; Scarrott et al. 1992; Lorenzetti et al. 2002b) and {\it
outside} of it (in Taurus - M\'{e}nard \& Duch\^{e}ne 2004). 

Given its location, VMR-D represents a well suited case to test the influence of the Galactic magnetic field on the jet
orientation. Remarkably, out of 15 jets, 8 (\# 1, 4, 6, 8, 9, 11, 13, 15)
have a position angle of $\sim$ 160$^{\circ}
\pm$ 15$^{\circ}$ (see Table\,\ref{tab:tab1}).
For completeness, 
also the jet from IRS8 (located in VMR, as well)
presents an orientation of $\sim$ 153$^{\circ}$
(Lorenzetti et al. 2002b). Hence, the present survey suggests that the Galactic magnetic field may 
be important during the early phases of the pre-main sequence evolution.
However, a selection effect could bias our sample and make this inference 
weaker: since the influence of the Galactic magnetic field tends to decrease toward the inside of the cloud, we expect that
jets aligned with it are located preferentially in the external layers. Therefore, these jets are likely the least embedded and therefore
those most easily detected.

\subsection {Comparison with other H$_2$-jet surveys}\label{sec:sec5.5}

The present H$_2$-jet survey covers a rather small region (about 1.2 deg$^2$), but it has the advantage of being substantially unbiased and therefore complete at the {\it Spitzer} sensitivity. Hence it is worthwhile to compare our results with those stemming from other similar surveys that, having both investigated larger areas and detected a larger number of sources, offer a more robust statistical significance in tracing the early phases of the stellar formation. Indeed, the peculiarity of our survey is to study a region inside a GMC, which is the typical site of the star formation, but rather different from other nearer dark clouds where star formation  occurs in a usually less turbulent environment. 

There are currently two main projects regarding 2.12\,$\mu$m H$_2$ jet emission:  {\it (i)} the survey UWISH2 (UKIRT Widefield Infrared Survey of H$_2$, Froebrich et al. 2011); and {\it (ii)} MHOs (Molecular Hydrogen Emission-Line Objects\footnote{Available at:http://www.jach.hawaii.edu/UKIRT/MHCat/}), namely the general catalog of H$_2$ infrared (1-2.2 $\mu$m) emission from young outflowing objects (Davis et al. 2010). On both papers references are given to the numerous 2.12\,$\mu$m H$_2$ imaging surveys of different regions carried out in the past years. Since UWISH2 will eventually provide an unbiased and complete census of the 1-0S(1) ro-vibrational H$_2$ line emission along the Galactic Plane, beside star forming and HII regions, it will include also evolved (AGB) stars, supernova remnants and fluorescent H$_2$ emission. 
Conversely, MHOs catalog is a fundamental database only of YSOs jets and HH objects that present molecular emission. Some of the jets described in the present paper have been already mentioned in a very preliminary form (Giannini et al. 2007), and, as such, are included in MHOs. Finally, an example rather similar to our VMR-D survey (as far as its GMC nature, size of the investigated area, distance from the Sun are concerned) is represented by the extensive Orion A survey (Stanke et al. 2002). Some results from the literature are the following:

\begin{itemize}
\item[UWISH2] (a) A potential source candidate can be assigned to about half of the jets; (b) typically, the flows are clustered in group of 3-5 objects, within a radius of 5 pc; (c) these groups are separated by about 5 pc; (d) 2/3 of the surveyed area is devoid of jets; (e) a large range of apparent lengths is found (from 4 to 130 arcsec); (f) a small fraction (about 10\%) are "parsec scale" jet; (g) there is a marginal (less than 3\,$\sigma$) over-abundance of flow position angles roughly perpendicular to the Galactic Plane. All of these are preliminary considerations based on the Serpens and Aquila survey (Ioannidis \& Froebrich, 2012).
\item[MHOs] (a) in some regions multiple knots and bow shocks radiate in many directions from a tight cluster of young stars and the relationship between these objects is often unclear,
 mainly in massive star forming clusters; (b) jets are found in both low- and high-mass star forming regions.
\item[Orion A] (a) the first outcome of this survey is the large number of detected jets
distributed in a more isolated way with respect to Orion B that presents a more clustered distribution; (b) a large variety of morphologies have been detected, ranging from large, extended, filamentary features to compact or even unresolved knots. This circumstance makes the identification of flows not always an objective process; (c) for a reasonable fraction of the jets a candidate exciting source is either obvious or suggested; (d) few jets show some degree of symmetry; (e) very  few jets are well collimated, virtually unresolved orthogonally to the flow direction; (f) apparently, there is not a preferred flow orientation.
\end{itemize}

The results of our VMR-D survey differ in several respects from the previous surveys' findings, mainly because these are based only on near-IR data. In the majority of 
cases (13 out of 15) a convincing exciting source has been found: this circumstance is related to the increased capability offered by the longer wavelengths coverage to identify  young and 
embedded objects. Practically, all the VMR-D jets display a high degree of collimation and most of them are composed of knots arranged in symmetrical position. Preserving the collimation and a certain symmetry over a large part of the GMC is a key-feature that makes easier the identification of a candidate driving source
and it could signal the presence of a relatively quiet environment, not too much affected by (local) turbulence. In this sense, VMR-D could be different from both the more active VMR-A,-B,-C clouds and Orion GMC; indeed, this picture is also supported
by independent evidences as the lack in VMR-D of O-B associations (Liseau et al. 1992)
and the overall gas dynamics (Olmi et al. 2010).

\section{Summary}
This paper presents a systematic search of molecular jets from young protostars in the
star formation region Vela, cloud D. Our color-color analysis in the four bands of \emph{Spitzer}-IRAC has led to the following results:
\begin{itemize}
\item[-] 15 jets have been detected in an area of about 1.2 square degrees, as well as a few sparse knots not displaying a clear alignment. All except two jets 
were previously known since already imaged in the H$_2$ 2.12\,$\mu$m line. The absence of jets in the part of the IRAC map not covered also in the near-infrared
strongly indicates that jets can be preferentially found in proximity of the dust peaks, since these are the locations where the 2.12\,$\mu$m images have been taken.
\item[-] Using the available catalog of YSOs in VMR-D we searched the jet exciting sources. For each jet, we selected all nearby Class\,I, flat, and Class\,II  sources, and selected those at the jet center and aligned with it. In this way we are able to find a best-candidate exciting source in all except two jets, for which we remain with two possible alternatives. To validate the jet-ES association, we have investigated whether a differential extinction exists between the two (hypothetical) lobes. Through an analysis of $L_{\rm{H_2}}/L_{\rm{IRAC}}$, we find a significant extinction gradient of $\Delta$A$_V \ge$ 5 mag in five cases.
\item[-] A search for the ES counterparts at wavelengths between 1.2\,$\mu$m and 1.2\,mm has been done. All except one source have a 24\,$\mu$m counterpart, and 10 are
associated with a 1.2\,mm peak. Four sources are very barely or not observed at wavelengths shorter than 24\,$\mu$m, suggesting they are very young protostars; three of them are also associated with the most compact jets ($l \la$ 0.1 pc).
\item[-] The SEDs of the ES have been constructed based on all available photometric data between 1.2\,$\mu$m and 1.2\,mm. We used the Robitaille et al. (2006, 2007) model to derive the main source parameters, such as stellar temperature and mass, envelope mass and radius, and total bolometric luminosity. On average, the ES are low-mass and relatively evolved protostars, with the
possible exception of few objects located close to infrared clusters, which could be intermediate-mass protostars as well as binaries or multiple objects not resolved at the \emph{Spitzer} angular resolution.
\item[-] A significant correlation is found between the projected jet length and the [24] - [70] color, which is consistent with an evolutionary scenario according to which shorter jets are associated with younger sources. We remark, however, that this trend has to be considered just in a statistical sense, and that no conclusions can be drawn for the individual sources, given the strong assumptions (e.g. jet inclination) under which the correlation is derived. The same trend is not recognizable if the diagram is drawn with shorter wavelengths colors, whose values are significantly affected by geometrical and extinction effects. 
\item[-] A coarse correlation is found between IRAC line cooling and the ES bolometric luminosity, which is in agreement with previous literature determinations. It is suggested that the sources segregate in two distinct
groups, to which pertain a different efficiency in the line cooling. This trend is in agreement with the the scenario
in which mass loss and mass accretion are tightly related phenomena, with mass accretion rates decreasing with time.
\item[-] About half of the jets present a position angle which may indicate a certain influence by the Galactic magnetic field, at variance with other star forming regions not located exactly in the Galactic plane. However, a selection bias could affect our sample. Indeed, since the influence of the Galactic magnetic field tends to decrease toward the inside of the cloud, we expect that jets aligned with it are located preferentially in the external layers, where they are most easily detected.
\item[-] Unlike other similar studies conducted on other star forming regions, the association between jet and ES has been more successful in VMR-D. Together with the increased capabilities offered by the longer wavelengths at which VMR-D is observed, this circumstance is likely favored by the high degree of collimation displayed by the jets. This, in turn, reflects the presence of a relatively quiet environment, for example in comparison with the other Vela clouds where star formation is going on.
\end{itemize}

\section{Acknowledgements}
This paper is dedicated to the memory of Alberto Salama who prematurely
passed away during the development of the work. 

It is based on observations
made with the {\it Spitzer Space Telescope}, which is operated by the Jet
Propulsion Laboratory, California Institute of Technology under
a contract with NASA.

{}

\begin{appendix}
\section{Appendix : notes on individual jets}\label{sec:appendix}
Below, we present a short description of both the identified jets and their candidate exciting sources. The numbering is that given in Tables \ref{tab:tab1} and \ref{tab:tab3}. References to papers where possible associations are described (e.g. sub-mm cores, BLAST sources, near-IR imaging) are given only for {\it Jet 1} and no longer repeated.

\begin{itemize}
\item[-] {\it Jet 1} - Extended bipolar jet composed of eight relatively bright knots. Its morphology is reminiscent of a {\it S-}shape. Blue- and red-shifted jet components correspond to the NW (knots 1-2-3-4) and SE (knots 5-6-7-8) lobe, respectively. It is located in a region dominated by the strong diffuse emission from a compact HII region (263.619-0.533 - Caswell \& Haynes 1987) associated with an IR cluster (Massi et al. 2003) named IRS16 following the nomenclature by Liseau et al. (1992). The potential driving source emerges only in the IRAC bands up to 24\,$\mu$m showing a SED with a Class\,I spectral index. This source is also associated with a BLAST (Olmi et al. 2009) and a 1.2 mm dust core (Massi et al. 2007). It is not aligned with the sequence of identified knots, but only with few of them of the NW lobe ($\#$ 1,2,3,4). Therefore, we can not confirm this source as the jet ES. See Figure\,\ref{jet_1:fig}.
\item[-] {\it Jet 2} - One of the nicest examples of very compact (0.13 pc) bipolar jet with well defined blue- (NE) and red-shifted (SW) lobes bracketing a very young exciting source. This latter is point-like and detected only in the MIPS 24,70 $\mu$m bands without any IRAC counterpart.  Given its very red [24]-[70] color of 7.4, the system appears a suitable target to investigate the very early stages of the mass ejection (and accretion) process. An observational benefit is offered by its peculiar location rather isolated (see Figure\,\ref{all:fig}) and unperturbed by excessive turbulence occurring in proximity of HII regions or IR clusters. Remarkably, the source is associated with a BLAST source and with a compact dust core unresolved at 1.2\,mm. See Figure\,\ref{jet_2:fig}.
\item[-] {\it Jet 3} - The morphology is very similar to that described for jet 2: both
show surprisingly the same PA orthogonal to the Galactic magnetic field. It is composed of only two knots (the blue-shifted one at NE and red-shifted one at SW) aligned in a very compact (0.08 pc) and collimated shape. Both knots are symmetrical with respect to the position of the driving source which, again, is detected only in the MIPS bands with a [24]-[70] color of 6.0. Also this jet lies in a quiet portion of VMR-D. See Figure\,\ref{jet_3:fig}.
\item[-] {\it Jet 4} - It is one of the two longest (0.68 pc) jets identified in the present survey, composed of the brightest detected knots. It preserves a high degree of collimation along all its length and shows bow-shock morphologies in both lobes oriented in opposite directions. 
Blue- and red-shifted lobes (northward and southward, composed by knots 1,2,3,4 and 5,6 respectively) are emanated from a well identified IR cluster (IRS20), where the source $\#$98 (Massi et al. 1999) seems associated with a bipolar IR nebula oriented along the same jet axis.
This source also coincides with a BLAST source and a resolved 1.2\,mm dust core, with a spectral index typical of a Class\,I source. The present observations allow us to reconsider our early suggestion coming from a lower sensitivity
survey (Lorenzetti et al. 2002b), according to which the northern bow shock was associated with a faint (unrelated) star. See Figure\,\ref{jet_4:fig}.
\item[-] {\it Jet 5} - Jet composed of well aligned knots. The membership of knot 1 to this jet is however doubtful, because this knot  is around 170\,arcsec (i.e. 0.57 pc) distant from the others. 
Two objects appear as possible candidate ES : one (5a, to which we give higher priority) located between knots 2-6 and 7, in which case knot\,1 should be unrelated to jet\,5, and one (5b) between knot\,1 and all the others, in which case 
the jet should be parsec-scale. Source 5a is barely detected with IRAC, while it is bright at 24\,$\mu$m. Source 5b is detected at all the wavelengths and it is associated both with a BLAST source and with an unresolved dust core. It shows a SED with a flat spectral index, which possibly signals it is not a very young object. See Figure\,\ref{jet_5:fig}.
\item[-] {\it Jet 6} - Jet composed of five knots perfectly aligned with a candidate driving source north of them; this latter is classified 
as Class I object with a bright IRAC/MIPS counterpart coinciding with a 1.2\,mm peak of dust emission; it is also part of the young IR cluster IRS19 (Liseau et al. 1992). 
According to this scenario the jet morphology results to be one-sided with its counter-jet possibly propagating northward deep inside the cloud. See Figure\,\ref{jet_6:fig}.

\item[-] {\it Jet 7} - Jet discovered in this survey. It is composed of two IRAC knots well aligned with a candidate exciting source with a SED with flat spectral index  and having valid fluxes up to 70\,$\mu$m. The jet is apparently one-sided, displaying just a SE lobe. See Figure\,\ref{jet_7:fig}.

\item[-] {\it Jet 8} - Two knots aligned in positions symmetrically opposed with respect to the candidate ES located at the system center. According to our analysis the red-shifted (blue-) lobe  extends northward (southward). The source displays a Class\,I SED with valid detected counterparts from 1.6\,$\mu$m to 70\,$\mu$m; it is also associated with both a BLAST source and a resolved 1.2\,mm dust peak. See Figure\,\ref{jet_8:fig}.
\item[-] {\it Jet 9} - Compact jet composed of five knots not perfectly aligned with
the candidate ES. This is classified as a Class I source coinciding with a 1.2\,mm dust core. See Figure\,\ref{jet_9:fig}. 
\item[-] {\it Jet 10} - Tentatively considered as a bipolar jet emanated by a young object (whose SED is classified as Class\,I). It is associated with both a BLAST source and an unresolved dust peak. Such a candidate ES is located at the jet center, but slightly displaced with respect to the knot axis. For this reason, a different origin of the observed knots can not be ruled out. See Figure\,\ref{jet_10:fig}.
\item[-] {\it Jet 11} - It appears as the most compact jet of the present survey (0.07 pc), composed of two knots, being the southeastern the blue- and the northwestern the red-shifted one. They seem symmetrically positioned at both sides of a faint candidate driving source
detected only at 24\,$\mu$m and without any association with cores at longer wavelengths. Further observations are needed to better understand the true nature of this source. See Figure\,\ref{jet_11:fig}.
\item[-] {\it Jet 12} - Three knots very close to those of {\it Jet 11}, but presenting
a different alignment.  Blue- and red-shifted lobes are not straightforwardly recognizable, with some indication in favor of the blue (red)-shifted lobe displaced NE (SW). Their position along a well defined axis is compatible with a candidate ES located roughly in the central position. This source  is associated with both a BLAST and 1.2\,mm core. See Figure\,\ref{jet_12:fig}.
\item[-] {\it Jet 13} - Together with {\it Jet 4} it appears as the longest jet of our survey. It is composed of six detected knots, of which knot 3 is practically coincident with the candidate ES while knot 4 extends slightly displaced toward SW.  While knots 3-4 represent the most compact part of the jet,  pairs of knots 1-2 and 5-6 are located far away from the exciting sources, but still preserving the same good alignment. In particular, both pairs display a shape well associated to opposite bows: the overall jet morphology suggests the occurrence of two (at least) episodes of matter ejection and that the time elapsed between them has been rather long. 
At a good level of confidence, we consider the southern (northern) lobe as the blue-(red-)shifted one. The recognized exciting source is bright at 24\,$\mu$m, has an extremely red MIPS color ([24]-[70] = 7.9), and is associated with a BLAST and 1.2\,mm core. Although it is a suitable candidate of a genuine protostellar object in a very early evolutionary stage, some doubts exist about its recognition as the jet ES (at least as far as the two extreme groups of knots is concerned). Indeed, while the source appears as one of the coolest and youngest of our survey, the jet (tentatively associated with  it) is definitely  the longest: such an occurrence appears rather contradictory if we assume the [24]-[70] color as an age indicator. Indeed, if the jet were composed only of knots 3 and 4, its length should be only 0.11 pc. Therefore, we prefer to not consider this jet in the analysis of Figure\,\ref{fig:jetcolor}, which is based on the estimate of the projected jet length. See Figure\,\ref{jet_13:fig}.
\item[-] {\it Jet 14}  -  Jet composed of six knots roughly aligned in E-W direction. One of them (indicated as 14-1) could be alternatively attributed to the {\it Jet 15} (knot 15-5). Two different candidate ES have been found (labeled as 14a and 14b): the former (classified as a Class\,I, to which we give higher priority) belongs to an IR cluster and has a 2MASS and a MIPS counterpart, while the latter is a very faint IRAC source (without any WISE confirmation) with a flat spectral index. See Figure\,\ref{jet_14:fig}.

\item[-] {\it Jet 15}  - We have presented this jet along with the candidate ES in previous papers (Lorenzetti et. al 2002; Giannini et al. 2005) to which the reader is referred to have a more complete view of that region.  The correspondence of the knots identification given in those papers with that given here, is the following: the complex of knots previously identified as A1-A2-A3, corresponds here to 15-1, B1-B2 to 15-2, G1 to G5 to 15-3, I to 15-4. Knot 15-5 is identified here for the first time, but its connection with jet 15 or jet 14 is doubtful (see also {\it Jet 14}). In the present paper other sparse knots (K1 to K4) have been discovered in this region, but any association with a plausible ES cannot be provided.
See Figure\,\ref{jet_15:fig}.
\end{itemize}
\end{appendix}


\begin{deluxetable}{ccccc}
\tabletypesize{\tiny} 
\tablecaption{Jets found in VMR-D with IRAC. \label{tab:tab1}} 
\tablewidth{0pt}
\tablehead
{Jet Id.& Number of  & length            & P.A.           & References$^a$ \\
        & IRAC knots & ($^{\prime\prime}$)/(pc)     &($^\circ$)           &                }
\startdata
1       &  8        & 89/0.30            &  163           & 1,2            \\
2       &  4        & 34/0.13            &   57           & 1,2            \\
3       &  2        & 24/0.08            &   57           & 1              \\
4$^b$   &  6        & 200/0.68           &  174           & 1,2,3          \\
5$^c$   &  6(7)     & 106/0.36 (321/1.09)&   48 (54)      & 1,2            \\
6$^d$   &  5        & 74$^e$/0.25$^e$    &  178           & 1,2,3          \\ 
7       &  2        & 38/0.14            &  121           &  -             \\
8       &  2        & 112/0.37           &    5           & 2              \\
9       &  5        & 56/0.19            &    5           & 2              \\
10      &  2        & 106/0.36           &   43           & 2              \\ 
11      &  2        &  20/0.07           &   146          & 2              \\
12      &  3        &  86/0.29           &   112          & 2              \\
13$^c$  &  6(2)     & 206/0.70 (30/0.11) &   173          & 2              \\
14$^c$  &  6(5)     & 155/0.53 (76/0.26) &    94 (96)     &  -             \\ 
15$^{c,f}$  &  4(5)     & 128/0.44 (139/0.48)&  155 (156)     & 3,4            \\
\enddata
\tablenotetext{a}{~ 1- Giannini et al. 2007, 2- De Luca et al. 2007, 3- Lorenzetti et al. 2002, 4- Giannini et al. 2005. } 
\tablenotetext{b}{~ IRS20 in (3). }
\tablenotetext{c}{~ The number in parenthesis indicates that a different number of knots could belong to the jet. The jet properties are reported accordingly in the subsequent columns.}
\tablenotetext{d}{~ IRS19 in (3). } 
\tablenotetext{e}{~ Half size than quoted in (1), see text.}
\tablenotetext{f}{~ IRS17 in (4). } 
\end{deluxetable}

\begin{deluxetable}{ccccccccccc}																					
\tabletypesize{\tiny} 	
\rotate																				
\tablecaption{Jets photometry. \label{tab:tab2}}																					
\tablewidth{0pt} 																					
\tablehead{
Knot ID    & $\alpha$(J2000.0)$^a$&$\delta$(J2000.0)$^a$ & size$^b$  & F(2.12)$\pm$$\Delta$F(2.12)$^c$ & F(3.6)$\pm$$\Delta$F(3.6)$^c$ & F(4.5)$\pm$$\Delta$F(4.5)$^c$&F(5.8)$\pm$$\Delta$F(5.8)$^c$&F(8.0)$\pm$$\Delta$F(8.0)$^c$& L$_{\rm{IRAC}}$&  L$_{1-0S(1)}$ \\
	       &   (h:m:s)     &($\circ:\prime:\prime\prime$)& (arcsec$^2$)  & (mJy) &(mJy) &(mJy) &(mJy) &(mJy) & (10$^{-2}$ L$_{\odot}$)& (10$^{-2}$ L$_{\odot}$)}
\startdata      										        								     
1-1            &  08:45:33.8   & -43:51:24   &   15  & 0.171 $\pm$ 0.001	 &	 $<$0.3      &  2.73 $\pm$ 0.09  &	 $<$3	     &       $<$3    & 8.217  & 0.052   \\
1-2	       &  08:45:34.1   & -43:51:28   &   32  & 0.299 $\pm$ 0.002	 &	 $<$0.9      &  3.10 $\pm$ 0.09  &	 $<$1	     &       $<$1    & 9.330  & 0.091   \\
1-3	       &  08:45:34.2   & -43:51:41   &    9  & 0.130 $\pm$ 0.002	 &  0.7  $\pm$ 0.2   &  0.84 $\pm$ 0.04  &	 $<$3	     &       $<$3    & 2.740  & 0.040\\
1-4	       &  08:45:34.4   & -43:51:47   &   11  & 0.105 $\pm$ 0.002	 &	 $<$0.2      &  0.19 $\pm$ 0.06  &	 $<$2	     &       $<$3    & 0.557  & 0.032   \\
1-5	       &  08:45:35.2   & -43:52:00   &   24  & 0.074 $\pm$ 0.002	 &  0.52 $\pm$ 0.07  &  0.5  $\pm$ 0.1   &	 $<$3	     &       $<$4    & 1.739  & 0.023  \\
1-6	       &  08:45:35.4   & -43:52:12   &   28  & 0.185 $\pm$ 0.003	 &  0.98 $\pm$ 0.05  &  0.93 $\pm$ 0.03  &	 $<$2	     &       $<$6    & 3.076  & 0.056  \\
1-7	       &  08:45:35.7   & -43:52:22   &   14  & 0.090 $\pm$ 0.002	 &  0.48 $\pm$ 0.03  &  1.23 $\pm$ 0.04  &  3.8  $\pm$ 0.2   &       $<$2    &13.127  & 0.028  \\
1-8	       &  08:45:35.6   & -43:52:36   &   20  & 0.161 $\pm$ 0.002	 &  0.74 $\pm$ 0.03  &  1.17 $\pm$ 0.03  &  3.7  $\pm$ 0.3   &       $<$3    &12.797  & 0.049   \\
\cline{1-11}     	  	                        										  
2-1	       &  08:48:16.8   & -43:47:08   &   46  & 0.129 $\pm$ 0.002	 &  0.28 $\pm$ 0.01  &  0.66 $\pm$ 0.01  &  0.21 $\pm$ 0.03  &       $<$1    & 2.583  & 0.040  \\   
2-2	       &  08:48:16.0   & -43:47:14   &   15  & 0.025 $\pm$ 0.002	 &	 $<$0.03     &  0.31 $\pm$ 0.01  &	 $<$0.2      &       $<$3    & 0.001  & 0.008  \\
2-3	       &  08:48:15.4   & -43:47:17   &   12  & 0.034 $\pm$ 0.002	 &  0.12 $\pm$ 0.01  &  0.14 $\pm$ 0.01  &  0.05 $\pm$ 0.01  &       $<$2    & 0.561  & 0.010   \\
2-4	       &  08:48:14.2   & -43:47:26   &   20  & 0.066 $\pm$ 0.002	 &  0.14 $\pm$ 0.01  &  0.34 $\pm$ 0.01  &  0.93 $\pm$ 0.02  &       $<$1    & 3.332  & 0.020   \\
\cline{1-11}     	  	                        	     								       
3-1	       &  08:46:34.9   & -43:21:07   &   18  & 0.069 $\pm$ 0.002	 &  0.19 $\pm$ 0.01  &  0.76 $\pm$ 0.01  &  0.54 $\pm$ 0.02  &   0.7 $\pm$ 0.02 & 5.482  & 0.021\\
3-2	       &  08:46:34.0   & -43:21:16   &   21  & 0.055 $\pm$ 0.003	 &  0.22 $\pm$ 0.01  &  0.67 $\pm$ 0.04  &  1.20 $\pm$ 0.03  &   0.7 $\pm$ 0.04 & 6.953  & 0.017\\
\cline{1-11}     	  	                        										  
4-1	       &  08:49:25.8   & -43:15:41   &   15  & 0.076 $\pm$ 0.001	 &  0.19 $\pm$ 0.01  &  0.40 $\pm$ 0.01  &  0.4  $\pm$ 0.1   &       $<$0.6  &  2.239 & 0.023  \\
4-2	       &  08:49:24.8   & -43:15:50   &  120  & 0.839 $\pm$ 0.004	 &  2.95 $\pm$ 0.07  &  4.98 $\pm$ 0.03  &  8.2  $\pm$ 0.2   &       $<$2    & 35.798 & 0.256  \\
4-3	       &  08:49:25.6   & -43:15:54   &   60  & 0.894 $\pm$ 0.004	 &  2.65 $\pm$ 0.03  &  4.09 $\pm$ 0.04  &  5.7  $\pm$ 0.1   &       $<$1    & 26.866 & 0.273  \\
4-4	       &  08:49:25.4   & -43:16:17   &   85  & 0.984 $\pm$ 0.004	 &  2.39 $\pm$ 0.06  &  5.33 $\pm$ 0.07  &  15.7 $\pm$ 0.4   &       $<$3    & 55.056 & 0.301  \\
4-5	       &  08:49:26.6   & -43:18:02   &  370  & 1.044 $\pm$ 0.008	 &  1.3  $\pm$ 0.1   &  6.1  $\pm$ 0.1   &  22.0 $\pm$ 0.7   &       $<$6    & 72.709 & 0.319  \\
4-6	       &  08:49:27.2   & -43:18:58   &   70  & 0.165 $\pm$ 0.004	 &  0.35 $\pm$ 0.01  &  1.18 $\pm$ 0.01  &	 $<$0.3      &       $<$0.6  &  3.649 & 0.050   \\
\cline{1-11}     	  	                        										  
5-1	       &  08.48.42.4  & -43.29.43   &    12   & 0.040 $\pm$ 0.002	 &  0.12 $\pm$ 0.01  &  0.18 $\pm$ 0.01  &  0.26 $\pm$ 0.03  &  0.43 $\pm$ 0.03 &   2.369 & 0.012\\
5-2	       &  08:48:26.2  & -43:31:39   &    12   & 0.231 $\pm$ 0.003	 &  0.21 $\pm$ 0.01  &  0.95 $\pm$ 0.01  &  0.93 $\pm$ 0.08  &       $<$0.2     &   5.182 & 0.071\\
5-3	       &  08:48:25.4  & -43:31:44   &    44   & 0.085 $\pm$ 0.004	 &  0.08 $\pm$ 0.01  &  1.24 $\pm$ 0.02  &  0.13 $\pm$ 0.03  &       $<$0.3     &   4.151 & 0.026\\
5-4	       &  08:48:23.6  & -43:31:55   &    15   & 0.137 $\pm$ 0.003	 &  0.37 $\pm$ 0.01  &  0.70 $\pm$ 0.02  &  0.62 $\pm$ 0.06  &       $<$0.1     &   3.736 & 0.042\\
5-5	       &  08:48:22.7  & -43:32:04   &    24   & 0.214 $\pm$ 0.003	 &  0.63 $\pm$ 0.01  &  1.45 $\pm$ 0.01  &  2.41 $\pm$ 0.08  &  0.9  $\pm$ 0.2  &  12.710 & 0.065\\
5-6	       &  08:48:22.4  & -43:32:00   &     9   & 0.002 $\pm$ 0.002	 &  0.09 $\pm$ 0.01  &  0.13 $\pm$ 0.01  &	 $<$0.6      &       $<$0.1     &   0.405 & 0.001\\
5-7	       &  08:48:18.9  & -43:32:49   &    40   & 0.011 $\pm$ 0.003	 &  0.04 $\pm$ 0.01  &  0.15 $\pm$ 0.01  &	 $<$0.6      &       $<$0.3     &   0.462 & 0.004\\
\cline{1-11}     	  	                    	   								      				   
6-1	       &  08:48:53.2  & -43:31:10   &     9   & 0.027 $\pm$ 0.003	 &  0.04 $\pm$ 0.01  &  0.25 $\pm$ 0.01  &  0.37 $\pm$ 0.03  &       $<$2       &  1.661 & 0.008\\
6-2        &  08:48:53.3  & -43:31:24   &    12   & 0.034 $\pm$ 0.002	 &  0.16 $\pm$ 0.01  &  0.44 $\pm$ 0.09  &  0.37 $\pm$ 0.03  &       $<$2       &  2.276 & 0.010\\
6-3	       &  08:48:53.3  & -43:31:34   &    16   & 0.035 $\pm$ 0.002	 &  0.03 $\pm$ 0.01  &  0.14 $\pm$ 0.01  &	 $<$0.2      &	     $<$2       &  0.440 & 0.011\\
6-4	       &  08:48:53.3  & -43:31:49   &    22   & 0.104 $\pm$ 0.003	 &  0.52 $\pm$ 0.01  &  0.98 $\pm$ 0.02  &  0.46 $\pm$ 0.04  &       $<$3       &  4.216 & 0.032\\
6-5	       &  08.48.53.3  & -43:32:12   &    55   & 0.374 $\pm$ 0.003    &  0.95 $\pm$ 0.01  &  2.90 $\pm$ 0.03  &  4.2  $\pm$ 0.1   &  3.3  $\pm$ 0.1  & 27.648 & 0.114\\
\cline{1-11}     	  	                        										  
7-1	       &  08:48:45.9  & -43:16:14   &    24   &        -             &  0.09 $\pm$ 0.01  &  0.14 $\pm$ 0.01  &       $<$0.3      &       $<$0.3     &  0.434 &  -	 \\
7-2	       &  08:48:47.4  & -43:16:27   &    32   &        -             &  0.28 $\pm$ 0.01  &  0.65 $\pm$ 0.01  &  0.77 $\pm$ 0.03  &  0.6 $\pm$ 0.1   &  5.459 &  -	 \\
\cline{1-11}     	  	                        										  		 
8-1	       &  08:49:13.8  & -43:35:42   &    35   & 0.237 $\pm$ 0.003	 &  0.45 $\pm$ 0.01  &  1.31 $\pm$ 0.03  &  0.9  $\pm$ 0.2   &  0.6  $\pm$ 0.2  &  7.701 & 0.072\\
8-2	       &  08:49:13.0  & -43:37:22   &    51   & 0.030 $\pm$ 0.004	 &  0.41 $\pm$ 0.01  &  0.32 $\pm$ 0.01  &	$<$0.2      &	   $<$0.3       &  1.092 & 0.023\\
\cline{1-11}     	  	                        									  
9-1	       &  08:48:59.4  & -43:37:46   &    44   & 0.124 $\pm$ 0.004    &  0.42 $\pm$ 0.02  &  0.53 $\pm$ 0.01  &       $<$0.3      &       $<$0.2     &  1.722 & 0.038\\
9-2	       &  08:48:59.2  & -43:38:01   &    36   & 0.027 $\pm$ 0.004    &  0.14 $\pm$ 0.01  &  0.27 $\pm$ 0.01  &       $<$0.1      &       $<$0.2     &  0.837 & 0.008\\
9-3	       &  08:48:58.8  & -43:38:12   &    18   & 0.031 $\pm$ 0.003    &  0.02 $\pm$ 0.01  &  0.37 $\pm$ 0.01  &       $<$0.1	   &	   $<$0.2       &  1.127 & 0.010\\
9-4	       &  08:48:58.5  & -43:38:24   &    38   & 0.194 $\pm$ 0.003	 &  0.50 $\pm$ 0.02  &  2.10 $\pm$ 0.01  &  2.98 $\pm$ 0.08  &       $<$0.3     & 13.759 & 0.003\\
9-5	       &  08:48:58.6  & -43:38:33   &    22   & 0.095 $\pm$ 0.002	 &  0.45 $\pm$ 0.01  &  0.54 $\pm$ 0.01  &  0.41 $\pm$ 0.05  &       $<$0.3     &  2.788 & 0.029\\
\cline{1-11}     	  	                        										    
10-1           &  08:47:45.5  & -43:42:57   &    20   & 0.089 $\pm$ 0.001	 &  0.23 $\pm$ 0.01  &  0.69 $\pm$ 0.01  &  0.74 $\pm$ 0.03  &  0.6  $\pm$ 0.1  &  5.563 & 0.027\\
10-2           &  08:47:38.7  & -43:44:13   &    30   & 0.079 $\pm$ 0.002	 &  0.41 $\pm$ 0.01  &  0.38 $\pm$ 0.01  &  1.35 $\pm$ 0.03  &  0.3  $\pm$ 0.1  &  5.468 & 0.024\\
\cline{1-11}     	  	                        										    
11-1           &  08:46:56.4  & -43:52:42   &    13   & 0.021 $\pm$ 0.001	 &  0.18 $\pm$ 0.01  &  0.51 $\pm$ 0.01  &  0.42  $\pm$ 0.02 &	   $<$0.1       &  2.617 & 0.006\\
11-2           &  08:46:57.3  & -43:52:58   &    22   & 0.018 $\pm$ 0.003    &  0.04 $\pm$ 0.01  &  0.14 $\pm$ 0.01  &	$<$0.06      &	   $<$0.9       &  0.428 & 0.005\\
\cline{1-11}     	  	                        								    
12-1       &  08:46:47.3  & -43:52:44   &   35   & 0.137 $\pm$ 0.003    &  0.26 $\pm$ 0.01  &  0.74 $\pm$ 0.01  &  0.69  $\pm$ 0.02 &  0.37 $\pm$ 0.04 &  4.918 & 0.042\\
12-2       &  08:46:51.1  & -43:53:08   &    30   & 0.044 $\pm$ 0.003    &  0.14 $\pm$ 0.01  &  0.20 $\pm$ 0.01  &	   $<$0.6  &  0.32 $\pm$ 0.04   &  1.469 & 0.013\\
12-3       &  08:46:53.8  & -43:53:15   &    36   & 0.097 $\pm$ 0.003    &  0.34 $\pm$ 0.01  &  0.58 $\pm$ 0.01  &  0.62  $\pm$ 0.01 &  0.80 $\pm$ 0.04 &  5.440 & 0.030\\
\cline{1-11}     	  	                        										   
13-1$^d$   &  08:49:28.0  & -44:03:13   &    56   & 0.122 $\pm$ 0.004	 &  0.45 $\pm$ 0.01  &  1.04 $\pm$ 0.01  &  0.87  $\pm$ 0.03 &      $<$0.1      &  5.377 & 0.037\\
13-2$^d$   &  08:49:27.3  & -44:03:20   &    90   & 0.343 $\pm$ 0.004	 &  0.61 $\pm$ 0.01  &  1.36 $\pm$ 0.01  &  1.13  $\pm$ 0.04 &  0.50 $\pm$ 0.05 &  8.321 & 0.105\\
13-3$^e$   &  08:49:28.7  & -44:04:29   &    50   & 0.046 $\pm$ 0.004	 &  0.30 $\pm$ 0.01  &  1.19 $\pm$ 0.01  &  0.48  $\pm$ 0.03 &  1.09 $\pm$ 0.01 &  7.679 & 0.014\\
13-4       &  08:49:29.4  & -44:04:53   &    45  & 0.097 $\pm$ 0.004    &  0.34 $\pm$ 0.01  &  0.56 $\pm$ 0.08  &	     $<$0.1	&      $<$0.2   &  1.784 & 0.030\\
13-5$^d$   &  08:49:28.2  & -44:06:28   &    15   & 0.027 $\pm$ 0.004    &  0.11 $\pm$ 0.01  &  0.18 $\pm$ 0.01  &	     $<$0.1	&      $<$0.1	 &  0.572 & 0.008\\
13-6$^d$   &  08:49:29.5  & -44:06:34   &    20   & 0.018 $\pm$ 0.001    &  0.04 $\pm$ 0.01  &  0.14 $\pm$ 0.01  &	     $<$0.1	&      $<$0.1	 &  0.424 & 0.006\\
\cline{1-11}    
14-1$^f$   &  08:46:36.0  & -43:56:02   &    36   &       -	             &  0.13 $\pm$ 0.01  &  0.24 $\pm$ 0.01  &        $<$0.2     &   0.7 $\pm$ 0.3  &  2.672 &  -	 \\ 	
14-2	   &  08:46:28.7  & -43:55:57   &    24   &      -               &  0.10 $\pm$ 0.02  &  0.29 $\pm$ 0.01  &  0.63  $\pm$ 0.04 &	   $<$0.2       &  2.429 &   -    \\
14-3       &  08:46:27.0  & -43:55:52   &    30   &      -               &  0.27 $\pm$ 0.01  &  1.04 $\pm$ 0.01  &  0.73  $\pm$ 0.03 &  0.60 $\pm$ 0.06 &  6.578 &   -    \\
14-4	   &  08:46:24.3  & -43:55:45   &    15   &      -               &  0.05 $\pm$ 0.01  &  0.10 $\pm$ 0.01  &  0.14  $\pm$ 0.02 &	   $<$0.1       &  0.651 &   -    \\
14-5	   &  08:46:21.8  & -43:55:46   &    45   &      -               &  0.48 $\pm$ 0.01  &  1.10 $\pm$ 0.08  &  0.75  $\pm$ 0.04 &	   $<$0.3       &  5.285 &   -    \\
14-6	   &  08:46:21.7  & -43:55:56   &    10   &      -               &  0.07 $\pm$ 0.01  &  0.26 $\pm$ 0.01  &  0.11  $\pm$ 0.01 &  0.16 $\pm$ 0.02 &  1.512 &   -   \\
\cline{1-11}     	  	                         										   
15-1	   &  08:46:30.9  & -43:53:18   &    60  &       -	             &  0.81 $\pm$ 0.02  &  1.95 $\pm$ 0.01  &  2.6 $\pm$ 0.2	 &   6.8 $\pm$ 0.4  & 30.121 & 0.101 \\
15-2	   &  08:46:31.8  & -43:53:51   &    65  &       -	             &  0.42 $\pm$ 0.01  &  1.06 $\pm$ 0.01  &  1.5 $\pm$ 0.1	 &   0.5 $\pm$ 0.3  &  8.230 & 0.039 \\
15-3	   &  08:46:35.4  & -43:55:08   &    110  &       -	             &  1.82 $\pm$ 0.02  &  5.96 $\pm$ 0.03  &  2.5 $\pm$ 0.5	 &   9.9 $\pm$ 0.1  & 50.312 & 0.186 \\
15-4	   &  08:46:35.7  & -43:55:22   &    25  &       -	             &  0.10 $\pm$ 0.01  &  0.13 $\pm$ 0.02  &  0.7 $\pm$ 0.1	 &   1.1 $\pm$ 0.1  &  4.902 & 0.014 \\
15-5$^f$   &  08:46:36.0  & -43:56:02   &    36  &       -	             &  0.13 $\pm$ 0.01  &  0.24 $\pm$ 0.01  &      $<$0.2	 &   0.7 $\pm$ 0.3 	    &  2.672 &  -  \\ 
\cline{1-11}     	  	                        	  	 								   
K-1	       &  08:46:30.1  & -43:54:11   &    18   &      -               &  0.22 $\pm$ 0.01  &  0.46 $\pm$ 0.04  &  0.34  $\pm$ 0.05 &  1.07 $\pm$ 0.06 &  5.060 &   -   \\
K-2	       &  08:46:27.5  & -43:54:15   &    28   &      -               &  0.05 $\pm$ 0.01  &  0.14 $\pm$ 0.01  &       $<$0.06     &  0.18 $\pm$ 0.04 &  0.886 &   -   \\
K-3	       &  08:46:29.1  & -43:54:40   &    20   &      -               &  0.07 $\pm$ 0.01  &  0.26 $\pm$ 0.01  &  0.20  $\pm$ 0.04 &	   $<$0.1       &  1.291 &   -   \\
K-4	       &  08:46:29.7  & -43:54:49   &    19   &      -               &      $<$0.02	   &  0.10 $\pm$ 0.01    &  0.07  $\pm$ 0.02 &  0.35 $\pm$ 0.06   &  1.380 &   -   \\
\enddata     		  	                  	     	  										
\tablenotetext{a}{~ Center of the corresponding box.}    
\tablenotetext{b}{~ Evaluated on the image at 4.5\,$\mu$m.}
\tablenotetext{c}{~ Errors are at 1\,$\sigma$ level, while upper limits are at 3\,$\sigma$ level.}	
\tablenotetext{d}{~ Knots tentatively attributed to jet\,\#13.}	
\tablenotetext{e}{~ The photometry is contaminated by the emission of the exciting source.}	
\tablenotetext{f}{~ Knot that could belong either to jet 14 (knot 14-1) or to jet 15 (knot 15-5).}	
\end{deluxetable}									    					     					 

\begin{deluxetable}{cccccccccccccc}
\tabletypesize{\tiny} 
\rotate		
\tablecaption{Exciting sources: IRAC candidates and counterparts$^a$.
\label{tab:tab3}} \tablewidth{0pt}
\tablehead{Source$^b$  &  RA (J2000.0)$^c$ &Dec (J2000.0)$^c$          & \multicolumn{6}{c} {ID} & \multicolumn{5}{c} {Separation (arcsec)}\\
                       & (h m s)        &  ($\circ \prime \prime\prime$)         &   IRAC    &   2MASS           & WISE                & MIPS  &   BLAST             &  SIMBA$^d$   &  2MASS     &   WISE    &   MIPS    &  BLAST       &  SIMBA  }
\startdata
            1  & 08 45  34.5	 & -43 51  57.3    &   18064   &     -             & J084534.45-435157.4 & 141	 &  J084535-435156     &   MMS2    &   -         &  0.5      &  1.0      & 7.0          & 7.5    \\
            2  & 08 48  15.9	 & -43 47  15.6    &	 -     &     -             &	  -	             & 657	 &  J084815-434714     &   umms16  &   -         &  -        &  -        & 4.2          & 8.1    \\
            3  & 08 46  34.3	 & -43 21  11.2    &     -     &     -             &	  -	             & 298	 &  -	               &	 -     &   -         &  -        &  -        &  -           &  -      \\
            4  & 08 49  26.3	 & -43 17  10.8    &  126339   &  08492636-4317125 & J084926.28-431711.1 & 991	 & J084925-431710	   &   MMS22   &   1.9       &  0.4      &  1.6      &  5.0         & 12.3    \\ 
            5a & 08 48  20.5	 & -43 32  29.1    &   95972   &     -             &      -              & 678	 &  -	               &    -      &   -         &  -        &  1.1      &   -          &  -      \\
            5b & 08 48  33.9	 & -43 30  47.3    &  103109   &  08483394-433047  & J084833.95-433047.3 & 729	 & 	J084833-433056     &   umms19  &   0.5       &  0.6      &  0.6      &  9.2         & 9.3     \\
            6  & 08 48  53.2	 & -43 30  55.2    &  112469   &     -             & J084853.17-433055.0 & 838	 & -	               &   MMS16   &   -         &  1.9      &  2.1      &  -           & 5.9     \\
            7  & 08 48  44.4	 & -43 16  8.06    &  108306   &     -	           & J084844.38-431607.5 & 787	 &  -	               &	 -     &   -         &  0.5      & 1.3       &  -           &  -      \\
            8  & 08 49  13.0	 & -43 36  24.5    &  120963   &  08491298-4336257 & J084913.19-433623.3 & 928	 & J084912-433618      &   MMS21   &   1.2       &  2.4      & 1.8       & 9.2          &  4.2    \\
            9  & 08 48  58.7	 & -43 38  21.1    &  114979   &    -	           &	   -	         & 858	 & -	               &   MMS17   &    -        &  -        & 0.6       & -            & 12.0    \\
           10  & 08 47  42.8	 & -43 43  48.4    &   74548   &  08474283-4343482 & J084742.80-434348.3 & 546	 & J084742-434347	   &   umms11  &   0.4       & 0.1       & 0.4       & 5.6          &  9.8    \\
           11  & 08 46  56.8	 & -43 52  52.2    &     -     &     -             &	   -	         & 392	 &  -                  &	-      &    -        &  -        & 1.5       &  -           &   -     \\
           12  & 08 46  49.1	 & -43 52  54.3    &   48370   &     -             &	   -	         & 355	 & J084648-435257	   &   MMS5    &    -        &  -        &  0.9      & 6.2          &  14.3   \\
           13  & 08 49  28.7	 & -44 04  29.2    &     -     &     -             & J084928.69-440429.2 & 998	 & J084928-440426      &   MMS23   &    -        & 1.0       & 1.2       &7.1           &  10.5   \\
	       14a & 08 46  31.4     & -43 55  56.9    &   41256   &  08463136-4355569 & J084631.35-435557.1 & 285   &  -                  &    -      &   0.0       &  0.5      &  9.5      &  -           &  -      \\
	       14b & 08 46  25.3     & -43 55  44.8    &   38798   &     -             &       -             & -     &  -                  &    -      &   -         & -         &  -        &  -            &  -      \\
\cline{1-14} 
\enddata
\tablenotetext{a}{Source association is done with respect to the IRAC coordinates given in the first two columns. The searching radius is the sum of the positional uncertainties associated to each pair of instruments. These are: 2MASS:0.4$^{\prime\prime}$; WISE: 1.0$^{\prime\prime}$; MIPS: 6.0$^{\prime\prime}$ for sources detected at 24\,$\mu$m, 20.0$^{\prime\prime}$ for sources detected only at 70\,$\mu$m; BLAST: 8.0$^{\prime\prime}$; SIMBA: 13.0$^{\prime\prime}$.}
\tablenotetext{c}{For simplicity, the source name is the same as the jet. Priority is expressed in alphabetic order.}
\tablenotetext{c}{The source coordinates are from the IRAC catalog, apart from sources \# 2,3,11,13 whose coordinates are taken from the MIPS catalog.}
\tablenotetext{d}{Resolved and unresolved cores are labeled with the prefix MMS and umms, respectively (according to Massi et al. (2007).}
\end{deluxetable}

\begin{deluxetable}{ccccccccccccc}
\tabletypesize{\tiny} 
\rotate
\tablecaption{Photometry of exciting sources$^a$. \label{tab:tab4}} 
\tablewidth{0pt}
\tablehead
{ES   & J             & H            & K$_s$              &    IRAC1       &    IRAC2        &   IRAC3        &     IRAC4      &  WISE3        & MIPS24 (WISE4)$^b$          &   MIPS70          & SIMBA$^c$        &spect.index$^d$ \\
      &   1.23 $\mu$m &  1.66 $\mu$m &   2.16 $\mu$m      &    3.6 $\mu$m  &    4.5 $\mu$m   &   5.8 $\mu$m   &   8.0 $\mu$m   & 12 $\mu$m     &24 $\mu$m (22 $\mu$m)    &   70 $\mu$m       & 1200 $\mu$m  &             }
\startdata
1     & -             & -            & -                  & 3.8$\pm$0.2    & 16.1$\pm$0.7    &  40$\pm$1      &  67$\pm$2      & 128$\pm$7     &  905$\pm$87             &   $<$8886       & 3700         & 1.7        \\
2     & -             & -            & -                  &    -           &    -            &        -       &      -         &   -           & 26.8$\pm$0.7            &2469$\pm$36        &   60         & -          \\
3     &-              & -            & -                  &    -           &    -            &        -       &      -         &   -           & 12.7$\pm$0.2            & 343$\pm$17        &  -           & -          \\
4     & $<$1.0	      & $<$8.5       & 4.32$\pm$0.44      &   152$\pm$4    & 552$\pm$15      &  1496$\pm$30   &  2900$\pm$95   & 6221$\pm$80   & $>$4000 (34884$\pm$257) &48580$\pm$1566     & 1320         & 3.2        \\
5a    & -             & -            & -                  & 0.027$\pm$0.003& 0.072$\pm$0.004 & $<$0.3         &  $<$0.4        &   -           & 14.2$\pm$0.04           &  $<$180           &  -           & -          \\
5b    & 0.47$\pm$0.05 &1.48$\pm$0.12 &2.97$\pm$0.14       &   5.8$\pm$0.2  & 9.4$\pm$0.3     & 11.4$\pm$0.2   & 15.9$\pm$0.5   & 29$\pm$2      & 198$\pm$5               &1014$\pm$21        & 110          & 0.3        \\
6     & -             & -            & -                  & 2.35$\pm$0.07  & 13.5$\pm$0.4    & 41.2$\pm$0.9   &  70$\pm$1      & 86$\pm$2      & 1090$\pm$45             &4867$\pm$54        & 230          & 1.8        \\
7     & -             & -            & -                  & 0.73$\pm$0.03  & 1.94$\pm$0.08   &2.69$\pm$0.06   &2.87$\pm$0.05   & 4.5$\pm$0.2   & 90$\pm$3                &968$\pm$13         &   -          & 0.3        \\
8     &  $<$1.8       & 0.61$\pm$0.15& 2.63$\pm$0.31      & 5.3$\pm$0.5    & 11.0$\pm$0.7    &18.9$\pm$0.7    & 23.3$\pm$0.8   &  53$\pm$1     & 583$\pm$65              & 3648$\pm$232      & 1760         & 0.8        \\
9     & -             & -            & -                  &0.16$\pm$0.01   & 0.74$\pm$0.04   &1.12$\pm$0.04   &1.23$\pm$0.05   &    -          & 113$\pm$8               & 1175$\pm$56       &  760         & 1.3        \\
10    & $<$1.0        &$<$10.3       & 8.15$\pm$0.2       & 42$\pm$2       & 79$\pm$2        & 124$\pm$3      & 180$\pm$4      &  332$\pm$3    & 1163$\pm$80             & 3519$\pm$39       &  100         & 1.1        \\
11    & -             & -            & -                  &   -            & -               &   -            & -              &    -          & 6.2$\pm$0.3             & $<$144            &   -          & -          \\
12    & -             & -            &-                   & $<$ 0.2        & 0.22$\pm$0.01   &0.42$\pm$0.02   &0.85$\pm$0.03   &    -          & 32.9$\pm$1              & $<$390            &  280         & -          \\
13   & -             & -            & -                  &   -            &  -              & -              & -              &    -          & 4.9$\pm$0.5             & 811$\pm$19        &  140         & -          \\
14a   & 3.82$\pm$0.1  & 9.16$\pm$0.25& 12.7$\pm$0.38      & 9.7$\pm$0.3    & 9.8$\pm$0.5     & 11.1$\pm$0.5   & 30$\pm$2       &  271$\pm$4    & $>$4000 (5294$\pm$102)  & 11870$\pm$645     & -            & 0.7        \\
14b   & -             & -            & -                  & 0.38$\pm$0.01  & 0.40$\pm$0.01   &0.49$\pm$0.02   & 0.90$\pm$0.06  &  -            & $<12$                   & $<1476$           & -            & 0.2        \\
\cline{1-13}
\enddata
\tablenotetext{a}{Fluxes are given in mJy. Upper limits are at 3\,$\sigma$ level.}
\tablenotetext{b}{WISE band\,4 fluxes are given for sources saturated with MIPS at 70\,$\mu$m.}
\tablenotetext{c}{Photometric error of $\sim$ 20 \%, as estimated by Massi et al. (2007).}
\tablenotetext{d}{The spectral index is computed between 2.16\,$\mu$m and 12\,$\mu$m for sources with at least four valid fluxes.}
\end{deluxetable}


\begin{deluxetable}{cccc}
\tabletypesize{\tiny} 
\tablecaption{Differential extinction between jet lobes. \label{tab:tab5}} 
\tablewidth{0pt}
\tablehead
{Jet Id.& Blue knots$^a$ & Red knots$^a$            & $\Delta$A$_V$     \\
        &            &                      &(mag)              }
\startdata
1        &  1-2-3-4  &  5-6-7-8           &   6                 \\
2        &  1        &  4                 &   11                \\
3        &  1        &  2                 &   5                 \\
4        &  1-2-3-4  &  5                 &   6                 \\
5a$^b$   &  7        & 2-3-4-5-6          &   2                 \\
5b$^b$   & 2-3-4-5-6-7 &  1               &   2                 \\
8        &  2        & 1                  &   8                 \\
9        &  5        & 1-2-3              &   2                 \\
10       &  2        & 1                  &   0                 \\ 
11       &  2        & 1                  &   6                 \\
12       &  1        & 2                  &   4                 \\
13a$^c$  &  5-6      & 1-2                &   4                 \\
13a$^c$  &  4        & 3                  &  10                 \\ 
15     &  1-2       & 3-4                &    2          \\

\enddata
\tablenotetext{a}{~~Knots used to compute the differential extinction (see Table\,\ref{tab:tab2} and Figures\,\ref{jet_1:fig} to \ref{jet_13:fig}).} 
\tablenotetext{b}{~~For jet\,\#5, both the possibilities that sources a and b are ES are considered. If the ES is source 5a, the pairs of symmetric knots are 2-6 and 7 (in which case knot 1 does not belong to the jet), while if the ES is source 5b, the  pairs of symmetric knots are 1 and 2-7.} 
\tablenotetext{c}{~~For jet\,\#13, both the possibilities that knots 1-2/5-6 belong or not to this jet are considered.} 
\end{deluxetable}


\begin{deluxetable}{cccccccccc}
\tabletypesize{\tiny} 
\tablecaption{Source parameters$^a$. \label{tab:tab6}} 
\tablewidth{0pt}
\tablehead
{Source name & distance          & disk incl,   & A$_V$(inter)    &  A$_V$(circum)      & T$_\star$             & M$_\star$         & R$_{\rm{env}}$        & M$_{\rm{env}}$    & L	            \\
             &  (pc)             & (deg)        & (mag)           & (mag)               &  (K)                  &(M$_\odot$)        &  (10$^3$ AU)          &(M$_\odot$)        &(L$_\odot$)}
\startdata  
1            &  {\bf 1000} (758) & {\bf 20} (40)& {\bf 17} (21)   & {\bf 738} (1050)   & {\bf 3923} (3865)     & {\bf 0.87} (0.74) & {\bf 15.8} (13.2)     & {\bf 43} (47)     & {\bf 19} (13.7) \\
2$^b$        &  500-1000         & -            & -               &   -                &   51                  & -                 &   -                   &	-	              & 0.3-1.3	       \\
3$^b$        &  500-1000         & -            & -               &   -                &   47                  & -                 &   -                   &	-                 & 0.2-0.8	       \\
4            &  {\bf 692} (660)  & {\bf 55} (65)& {\bf 18} (24)   & {\bf 107} (876)    & {\bf 4162} (4162)     & {\bf 1.98} (1.98) & {\bf 2.6} (2.6)       & {\bf 0.08} (0.07) & {\bf 216} (216) \\
5b           &  {\bf 759} (832)  & {\bf 20} (20)& {\bf 11} (7)    & {\bf 23} (24)      & {\bf 2887} (3003)     & {\bf 0.15} (0.18) & {\bf 1.3} (3.0)       & {\bf 0.05} (0.14) & {\bf 1.1} (1.6) \\
6            &  {\bf 692} (912)  & {\bf 50} (40)& {\bf 61} (71)   & {\bf 37} (28)      & {\bf 3535} (3160)     & {\bf 0.40} (0.23) & {\bf 6.9} (2.2)       & {\bf 0.16} (0.03) & {\bf 16} (30)	 \\
7            &  {\bf 759} (501)  & {\bf 30} (50)& {\bf 43} (45)   & {\bf 26} (25)      & {\bf 3039} (3283)     & {\bf 0.20} (0.27) & {\bf 1.1}$^c$         & {\bf 0.05}$^c$    & {\bf 1.4} (0.8) \\
8            &  {\bf 759} (692)  & {\bf 20} (20)& {\bf 24} (21)   & {\bf 22} (80)      & {\bf 4370} (3949)     & {\bf 0.35} (0.68) & {\bf 7.0} (9.1)       & {\bf 3.5}  (8.3)  & {\bf 4.7} (5.6) \\
9            &  {\bf 692} (1000) & {\bf 20} (20)& {\bf 57} (67)   & {\bf 94} (2)       & {\bf 3040} (3620)     & {\bf 0.19} (0.42) & {\bf 3.3} (3.8)       & {\bf 2.7}  (3.5)  & {\bf 1.3} (2.8) \\
10           &  {\bf 955} (871)  & {\bf 20} (40)& {\bf 16} (7)    & {\bf 8}  (29)      & {\bf 2970} (3787)     & {\bf 0.18} (0.65) & {\bf 3.8} (7.9)       & {\bf 0.2} (0.3)	 & {\bf 12} (14)	 \\
13$^b$       &  500-1000         & -            & -               &  -                 &   43                  & -                 &   -                   &	-	             & 0.2-0.9	     \\
14a          &  {\bf 955} (661)  & {\bf 90} (90)& {\bf 10} (5)    & {\bf$>$ 1000} ($>$ 1000) & {\bf 4386} (4503) & {\bf 3.17}(2.63) & {\bf 5.7}$^c$        & {\bf 0.18}$^c$      & {\bf 109} (38)  \\
\enddata
\tablenotetext{a}{~~For each parameter, the value given in boldface refers to the best-fit model, while the value in parenthesis refers to the model whose  $\chi^2$ is 50\% higher than the minimum $\chi^2$.} 
\tablenotetext{b}{~~SED fitted with a grey-body function, by assuming a distance between 500 pc and 1000 pc. For these sources the fitted temperature is that of the envelope.} 
\tablenotetext{c}{~~uncertain since the SED is measured up to $\lambda$=70\,$\mu$m.}
\end{deluxetable}

\begin{figure}
\includegraphics[angle=0,width=12cm]{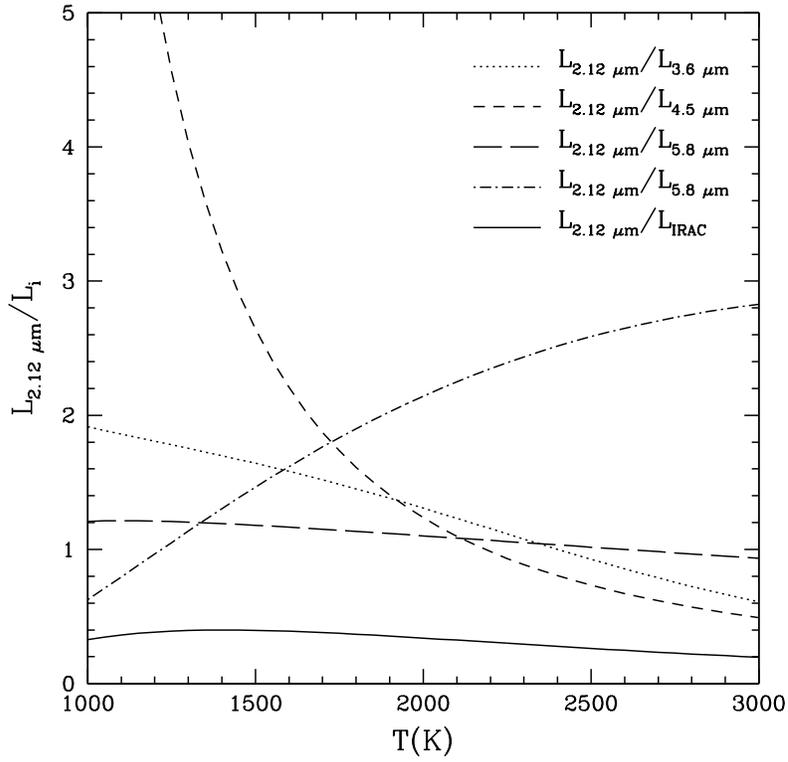} 
\caption{ Ratio of $L_{1-0S(1)}$ over IRAC luminosity in different bands  vs. gas temperature. LTE approximation is adopted.\label{LTE:fig}}
\end{figure}
\begin{figure}
\includegraphics[angle=0,width=12cm]{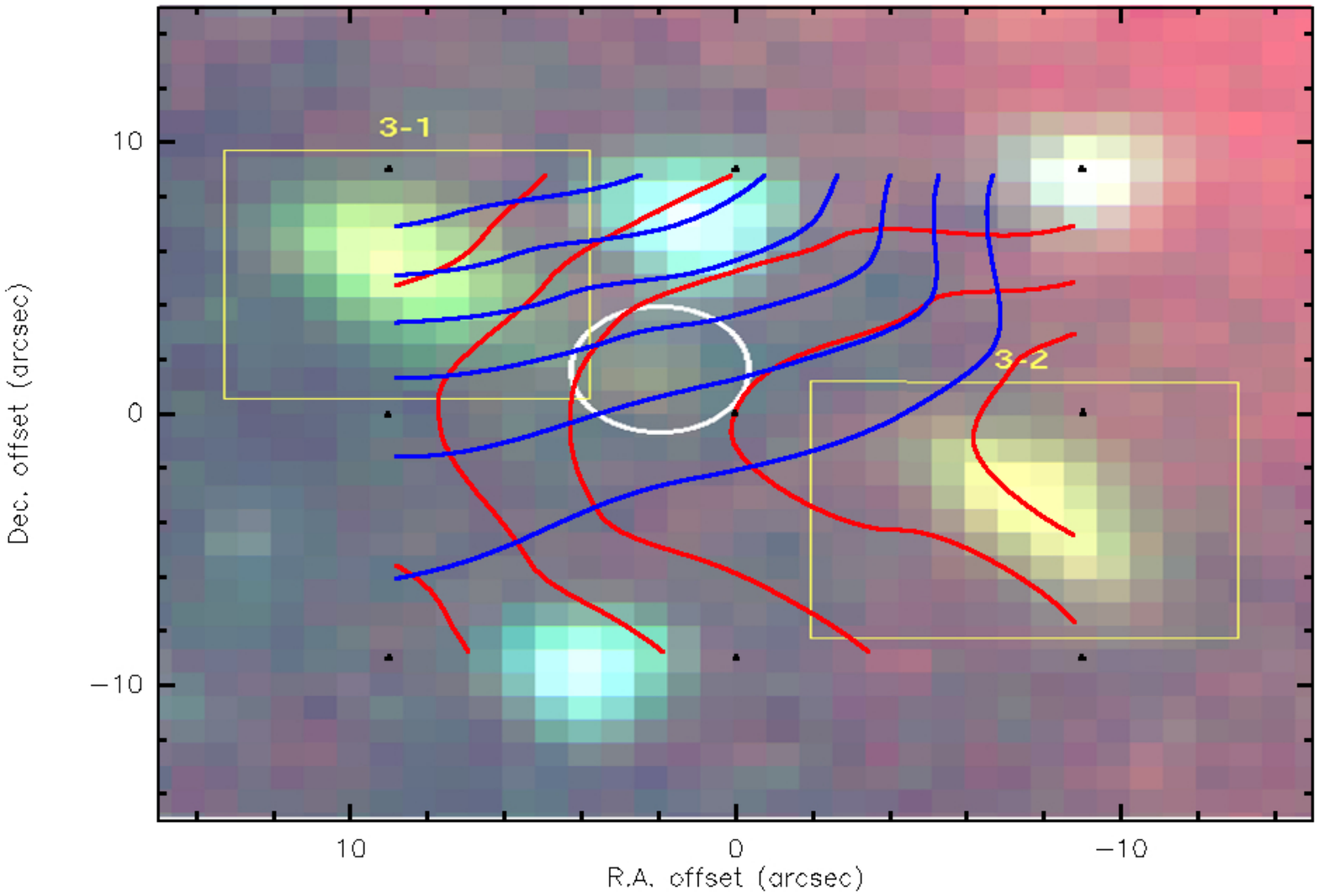}
\caption{$^{12}$CO(3-2) map of jet\,\# 3 superposed on the IRAC three-color image. Contours are (from right to left) 1.2, 1,8, 2.4, 3.0 and 3.6 K km s$^{-1}$ for the red-lobe and 0.6, 1.2, 1,8, 2.4, 3.0 and 3.6 K km s$^{-1}$ (from left to right) for the blue-lobe, respectively. The velocity is integrated between  -3.6 km s$^{-1}$ and +0.8 km s$^{-1}$ for the blue-wing, and 
between +0.8 km s$^{-1}$ and +3.4 km s$^{-1}$ for the right-wing. The positions of the exciting source and the H$_2$ knots are evidenced as well (ellipse and boxes, respectively).\label{fig:apex}}
\end{figure}

\begin{figure}
\includegraphics[angle=0,width=17cm]{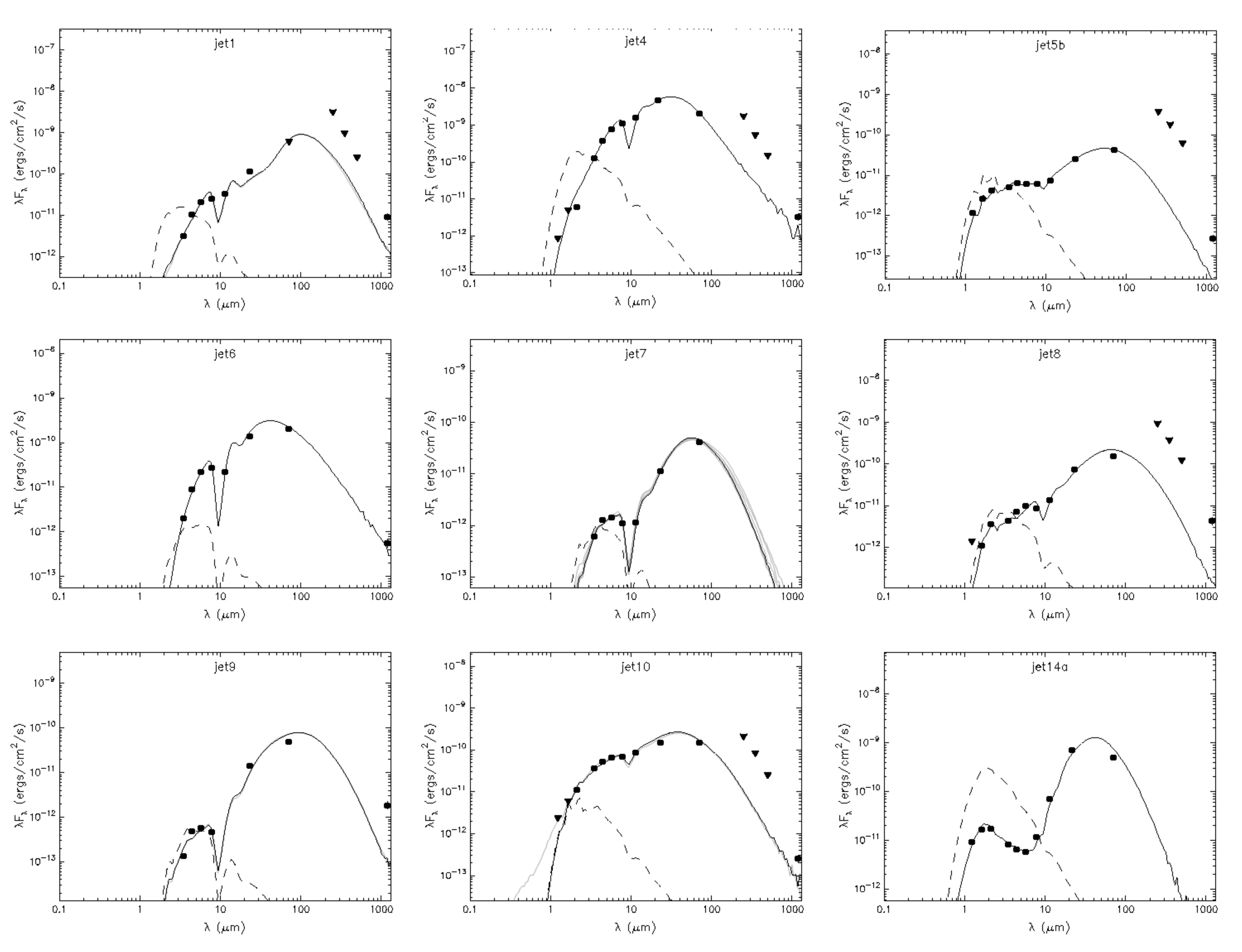}
\caption{Fits with the Robitaille et al. (2007) model of the SED of the ES. The filled circles show the observed fluxes, while triangles are upper limits. BLAST data points (represented as upper limits) are not included in the fit.
The black line shows the best-fit, and the gray lines show subsequent good fits. The dashed line shows the stellar photosphere corresponding to the central source of the best-fit model, as it would appear in the absence of circumstellar dust (but including interstellar extinction).\label{seds:fig}}
\end{figure}

\begin{figure} 
\includegraphics[angle=0,width=12cm]{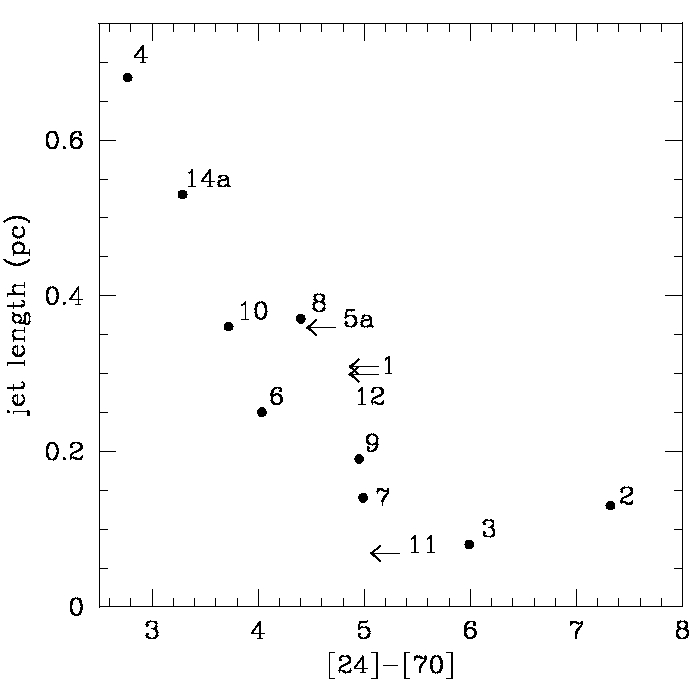}
\caption{Projected jet length (in pc) vs. the [24]-[70] color. Arrows are upper limits.\label{fig:jetcolor}}
\end{figure}

\begin{figure}
\includegraphics[angle=0,width=12cm]{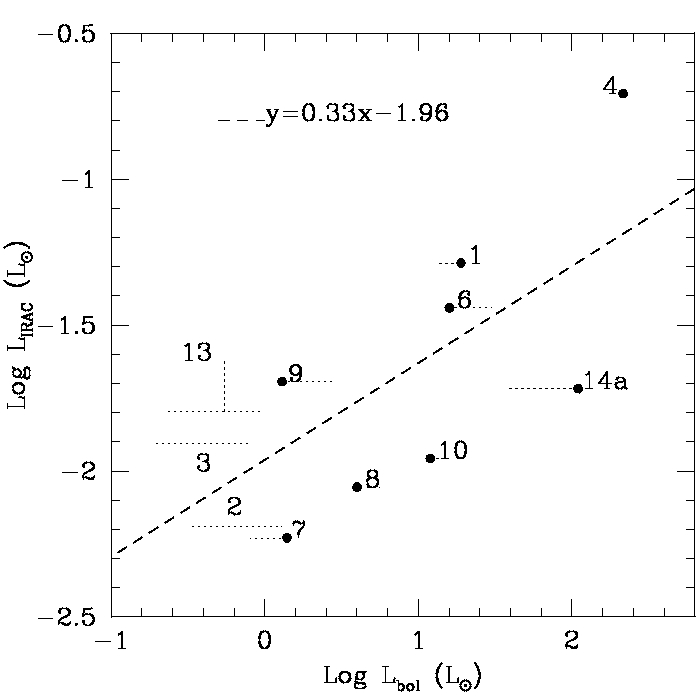}
\caption{IRAC line cooling vs. ES bolometric luminosity. The equation of the best-fit through the data points is given in 
the top side of the plot. Horizontal dotted lines indicate the range of bolometric luminosity given in Table\,\ref{tab:tab6}; in particular the value
associated with the best-fit model is indicated with a dot. For source \#\,13, the vertical line indicates $L_{\rm{IRAC}}$ in case the jet is composed alternatively of two or
six knots (see Appendix\,\ref{sec:appendix}).\label{fig:lumlum}}
\end{figure}

\end{document}